\documentclass[a4paper,11pt]{article}

\usepackage[utf8]{inputenc}
\usepackage[T1]{fontenc}
\usepackage{lmodern}
\usepackage[english]{babel}
\usepackage{geometry}
\usepackage{amsmath,amssymb,amsthm,mathtools}
\usepackage{bm}
\usepackage{booktabs,array,tabularx}
\usepackage{multirow}
\usepackage{subfigure}
\usepackage{algorithm}
\usepackage{algpseudocode}
\usepackage{float}
\usepackage{graphicx}
\usepackage{xcolor}
\usepackage[authoryear,round]{natbib}
\usepackage{hyperref}

\bibpunct{(}{)}{;}{a}{,}{,}

\hypersetup{
  pdftitle={Cure-Fraction Random Survival Forests: Separating Cure from Latency},
  pdfauthor={Pedro Luiz Ramos, Lucas Osses, Debajyoti Sinha},
  pdfkeywords={cure models, survival trees, random survival forests, long-term survivors, censored data},
  hypertexnames=false,
  hidelinks
}

\newcommand{\R}{\mathbb{R}}

\newcommand{\I}{\mathbb{I}}
\newcommand{\logit}{\operatorname{logit}}

\theoremstyle{plain}
\newtheorem{theorem}{Theorem}[section]
\newtheorem{lemma}[theorem]{Lemma}
\newtheorem{corollary}[theorem]{Corollary}
\newtheorem{proposition}[theorem]{Proposition}

\theoremstyle{definition}

\newtheorem{assumption}[theorem]{Assumption}

\theoremstyle{remark}

\begin{document}

\title{\Large\bfseries Cure-Fraction Random Survival Forests: Separating Cure from Latency}

\author{
Pedro Luiz Ramos$^{1,*}$,
Lucas Osses$^{1}$,
Debajyoti Sinha$^{2}$\\[4pt]
\small $^{1}$Facultad de Matem\'aticas, Pontificia Universidad Cat\'olica de Chile,\\[-1pt]
\small Avda. Vicu\~na Mackenna 4860, Macul, Santiago 7820436, Chile\\
\small $^{2}$Department of Statistics, Florida State University,\\[-1pt]
\small Tallahassee, FL 32306-4330, USA\\
\small $^{*}$Corresponding author: \texttt{pedro.ramos@uc.cl}
}

\date{}

\maketitle

\begin{abstract}
In many cancer and clinical studies, the population under study has a positive fraction of long-term survivors, motivating survival models that distinguish probability of long-term survival (cure fraction) from the latency of the subgroup that is susceptible to the event. Existing survival trees and forests partition data using criteria that emphasize differences in the ranks of survival times and therefore need not identify covariates associated with cure probability. 
We introduce  cure-fraction based survival trees (CBST) that recursively partition the covariate space using contrasts in tail survival probability supported by adequate follow-up. 
The resulting ensemble forest targets long-term survival heterogeneity directly rather than ranking of the survival times.
We establish consistency for the induced tail-survival target and derive conditions under which our procedure consistently recovers cure-rate related partitions and cure probabilities. The analysis clarifies the separate roles of follow-up maturity, localization, and residual-tail bias in identifying cure effects and supporting statistical inference. We further develop variance estimation and asymptotic results and uncertainty quantification for predictions from the ensemble forest.
Simulations show that the cure-fraction based splitting criterion can improve recovery
of cure heterogeneity relative to a log-rank based forest when cure incidence and latency are driven by different covariates. 
Analyses of 
temporally held-out cohorts from the S\~ao Paulo Hospital Cancer Registry
demonstrate that our method produces stable long-term survival risk groups while preserving substantial late follow-up support.
The proposed framework provides a nonparametric approach to learning and interpreting cure heterogeneity in settings where long-term survival is the primary scientific objective.
\end{abstract}

\vspace{0.5em}
\vspace{1em}

\section{Introduction}
\label{sec:introduction_new}

Persistent survival plateaus in cancer and other clinical studies motivated the
development of cure-rate models \citep{boag1949,berkson1952}. 
In such studies, two distinct sources of survival heterogeneity are of scientific interest: heterogeneity in the probability of long-term survival and heterogeneity in survival times among susceptible subjects. 
Semiparametric and parametric mixture cure models formalize this distinction by separating the cure probability \(\pi(x)\) and
the susceptible survival function $S_u(t\mid x)$ 
\citep{farewell1982,maller1996} given covariates $x$. Regression formulations allow $x$ to
have different effects on incidence and latency \citep{kuk1992,sy2000,peng2000}.
Conventional survival trees and forests, however, are constructed to separate survival distributions and therefore need not identify covariates that drive heterogeneity in probability of cure.
This distinction has a direct consequence for recursive partitioning. 

A covariate that strongly affects survival time of only a susceptible patient can yield a large log-rank statistic even when it has little effect on the cure probability. Conversely, a covariate that primarily affects the cure probability may produce relatively little separation in observed survival times. Thus, a survival forest optimized for observed survival time separation need not recover the partition relevant to long-term survival \citep{LeBlancCrowley1993,ishwaran2008}.
Both partitions may be
useful for predictions, but they address different scientific questions. Conventional discrimination measures may not expose
this difference because they reward correct survival-time ranking rather than
recovery of the long-term survivor structure \citep{ezquerro2023}.
Survival trees can represent nonlinear effects, interactions, and thresholds
without a prespecified linear predictor \citep{BreimanEtAl1984CART}.
Bagging and random-variable selection reduce the instability of individual
trees and provide the basis for random forests
\citep{Breiman1996Bagging,breiman2001}. However, taking a late survival
prediction from a log-rank forest does not change its splitting criterion.
Its terminal nodes may also contain too little late follow-up to support a
cure interpretation. If the follow-up is inadequate, a survival plateau alone does not establish the existence
of a cured fraction 
\citep{EscobarBachVanKeilegom2019,YuenMusta2026}.

We propose \emph{cure-fraction based survival trees} (CBST) and their random-forest ensembles whose splits are driven by differences in stabilized tail-survival probabilities rather than by survival-time separation. Because an estimated survival at a time $t$ represents the cure probability only when residual survival at that time $t$ for a susceptible subject is negligible, we incorporate explicit maturity constraints based on late risk-set and survival support.

The central distinction is between \emph{tail-survival}, \(\psi_n(x)\),
which summarizes survival over the observed tail, and \emph{cure probability},
\(\pi(x)\). They differ by the residual contribution of susceptible subjects:
\(\psi_n(x)=\pi(x)+\beta_n(x)\). A tail-survival summary from a conventional
survival forest need not estimate cure probability, even if it estimates
tail-survival accurately. CBST makes this distinction explicit in its splitting
target and follow-up requirements. Our results establish consistency for
tail-survival; consistency for cure probability additionally requires
\(\beta_n(x)\to0\), alongside the stated node-estimation and localization
conditions.

Our extensive simulation studies show that our splitting method improves recovery of the heterogeneity in cure probability when the cure probability and the latency of any susceptible subject are driven by different covariates, while also demonstrating the effects of follow-up maturity and localization bias on inference. Our analysis of temporally held-out data from the São Paulo Hospital Cancer Registry illustrates how maturity constraints can preserve long-term separation while reducing reliance on immature terminal nodes. 
The application also evaluates whether the resulting groups show distinct
tail-survival patterns with enough follow-up to estimate their tails directly.
The remainder of the paper is organized as follows. Section~2 introduces the
notation and cure-model decomposition. Sections~3 and~4 present the trees,
forests, and theoretical results. Sections~5 and~6 contain the simulation
study and FOSP application, respectively. Appendix~A gives the technical proofs.

Figure~\ref{fig:cure_vs_logrankprop} illustrates this distinction: the
CBST selects the plateau variable \(X_1\), whereas the log-rank tree selects the latency variable
\(X_2\). The corresponding true cure contrasts are 0.214 and 0.002.
The example illustrates the splitting targets rather than providing
a performance comparison.
Averaging these trees yields a random forest of CBST. Its direct
target is tail-survival \(\psi_n(x)\) supported by observed
follow-up; this approximates \(\pi(x)\) when residual susceptible survival is
negligible. Population survival, susceptible survival, and dynamic cure
probability are obtained as coherent secondary outputs under the conditions
stated below. Honest subsampling provides a framework for forest estimation
and inference \citep{WagerAthey2018,AtheyTibshiraniWager2019}, and
jackknife and infinitesimal-jackknife methods motivate the variance calculations
\citep{WagerHastieEfron2014,MentchHooker2016}.

The proposal builds on existing approaches to long-term survival.
Promotion-time models describe tumor latency through latent causes
\citep{yakovlev1996}, while nonmixture formulations represent long-term
survival without an explicit two-component mixture
\citep{tsodikov2003,rodrigues2009}. Bayesian constructions provide another
approach to incidence and latency \citep{chen1999}. Semiparametric mixture
models retain a regression structure for cure while relaxing the susceptible
event-time distribution \citep{kuk1992,sy2000,peng2000}. These approaches are
effective when the specified incidence and latency structures adequately
represent the covariate effects.

Nonparametric procedures further reduce functional-form restrictions
\citep{AmicoVanKeilegom2018,LopezChedaEtAl2021Npcure}, with large-sample and
pseudo-observation methods extending estimation to flexible regression
settings \citep{HanEtAl2024,ChiouEtAl2025}. Across these approaches, adequate
follow-up remains necessary to distinguish cure from very late susceptible
failures. This issue has motivated estimators for insufficient follow-up and
formal assessments of follow-up adequacy
\citep{EscobarBachVanKeilegom2019,YuenMusta2026}.

Recent machine-learning approaches include prediction-driven and
support-vector methods \citep{jiang2019,pal2023}, decision trees combined with
mixture-cure estimation and comparisons of alternative learners
\citep{aselisewine2023,aselisewine2024}, and Bayesian additive regression trees
\citep{LineroRubioBasak2026BartCure}. In the proposed method, cure
heterogeneity guides recursive partitioning, with follow-up maturity explicitly
included in the construction. An inverse probability of censoring weighted
(IPCW) node estimator is also considered when censoring depends on observed
covariates. This distinction is relevant even under conditional independence
of event and censoring times given the covariates, since independence need not
be preserved after subjects are pooled within a node
\citep{OvergaardEtAl2019,CuiEtAl2022}.

Our contribution has following components.
First, we introduce a novel splitting criterion of the covariate space that directly targets heterogeneity in the cure probabilities.
Second, we establish split-recovery and consistency results that distinguish estimation of risk of event from identification of cure probability.
Third, we develop an inferential framework that clarifies the role of follow-up maturity and residual-tail bias in our random forest-based analysis of survival data with cure rate.
We establish conditions for recovery of the tail-survival split and consistency
of the resulting forest under response honesty, uniform node estimation, and
localization. 
Our results clarify what can be estimated from the available follow-up and what additional
conditions are needed for inference.
Our simulation study evaluates cure-probability estimation, split recovery, and
uncertainty quantification. The methodology is illustrated using the
S\~ao Paulo Hospital Cancer Registry (FOSP), comparing long-term risk groups
and tail support in a later temporal cohort. 

\section{Notation and preliminaries}
\label{sec:notation}

Let $(\Omega, \mathcal{F}, \mathbb{P})$ be an underlying probability space. To formalize the distinction between long-term survivorship and survival time, we adopt the standard two-component mixture cure model framework (\cite{berkson1952})  where a random, unobservable indicator $Z \in \{0, 1\}$ dictates susceptibility to an event of interest: subjects with $Z = 1$ are non-susceptible to the event (cured). Let $T \in \mathbb{R}^+ \cup \{\infty\}$ denote the true survival time, satisfying $\mathbb{P}(T = \infty \mid Z = 1) = 1$. We observe independent right-censored data
$\{Y_i,\Delta_i,X_i\}$ for $ i=1,\ldots,n$,
where $Y_i=\min(T_i,C_i)$ is the bounded survival horizon, \(\Delta_i=\I(T_i\leq C_i)\) is the censoring indicator, and 
$X_i\in\mathcal{X}\subseteq\R^p$ is a vector of baseline covariates. We
assume the censoring time $C_i \in \mathbb{R}^+$ to be conditionally independent, \(T_i\perp C_i\mid X_i\), and have positive support, 
$P[C>t\mid X]>0$, for all $t$ in the time region $(0,\tau]$ of interest, and for all values of $X$.
For a subject with observed  $\Delta_i=1$, the true status is known to be  $\{Z_i = 0\}$, but for a censored subject with $\Delta_i=0$, the status of $Z_i$ is not known.
Given the covariate $X=x$, the population survival function is represented as
\begin{equation}
S(t\mid x)=P(T>t\mid X=x)
=\pi(x)+\{1-\pi(x)\}S_u(t\mid x),
\label{eq:mixture_model_new}
\end{equation}
where
$\pi(x)=P(Z=1\mid X=x)$, and $S_u(t\mid x)=P(T>t\mid Z=0,X=x)$
with \(S_u(t\mid x)\to 0\) to ensure that the limiting survival probability is
\(\lim_{t\to \infty}S(t\mid x)=S(\infty\mid x)=\pi(x)\). For us, the cure probability
\(\pi(x)\) is a primary estimand.
For a censored observation, cure status $Z_i$ is unobserved. Using Bayes' rule with Equation~\eqref{eq:mixture_model_new}, 
 the dynamic posterior cure probability of a patient remaining event-free at time \(y\) is
\begin{equation}
q(y,x)
=P(Z=1\mid T>y,X=x)
=
\frac{\pi(x)}
{\pi(x)+\{1-\pi(x)\}S_u(y\mid x)}\ .
\label{eq:dynamic_cure_new}
\end{equation}
 For an individual who remains event-free until time $y$,  \(q(y, x)\) is the conditional probability of being cured, $\{Z_i=1\}$, based on accumulated history at follow-up time $y$. This distinction is crucial because two distinct covariate profiles can yield identical fixed-horizon survival probabilities while corresponding to entirely different underlying cure fractions. The cure-fraction based splitting criterion targets the probability of long-term survival rather than these transient survival differences; \(q(y,x)\) is a derived prediction, not the quantity used to select splits.

Let \(A\subseteq\mathcal{X}\) denote an arbitrary subset (node) generated during recursive partitioning. We define the conditional node-level cure probability as
$\pi_A=P(Z=1\mid X\in A)$,
and the node-level survival probability of a susceptible subject as 
$S_{u,A}(t)=P(T>t\mid Z=0,X\in A)$,  with the corresponding averaged population survival function within node \(A\) is given by \(S_A(t) = \pi_A + \{1 - \pi_A\}S_{u,A}(t)\).
One popular example of the estimate \(\widehat S_A(t)\) of $S_A(t)$ is
the Kaplan--Meier estimator \citep{KaplanMeier1958}, however, later we also consider the IPCW-based estimator to adjust for covariate-dependent censoring. Existing survival tree methods (e.g., \cite{LeBlancCrowley1993, ishwaran2008}) produce node-specific $\widehat{S}_A(t)$ only as terminal-node summaries, while they grow nodes by maximizing hazard separation via rank-based statistics (such as log-rank). Such criteria collapse variations in incidence (\(\pi _{A}\)) and latency (\(S_{u,A}\)) into a singular metric, favoring covariates that alter short-term timing over those defining long-term survival plateaus.
To isolate long-term survival heterogeneity directly, we propose integrating $\widehat{S}_A(t)$ directly into the tree-growth phase to estimate a tail-survival functional. This functional corresponds to $\pi _{A}$ only under the separate follow-up condition, which eliminates residual survival among susceptible patients.
Consequently, our recursive splitting searches for daughters
with different \(\widehat S_A(t)\).
\begin{figure}[!h]
\centering
\includegraphics[width=\textwidth]{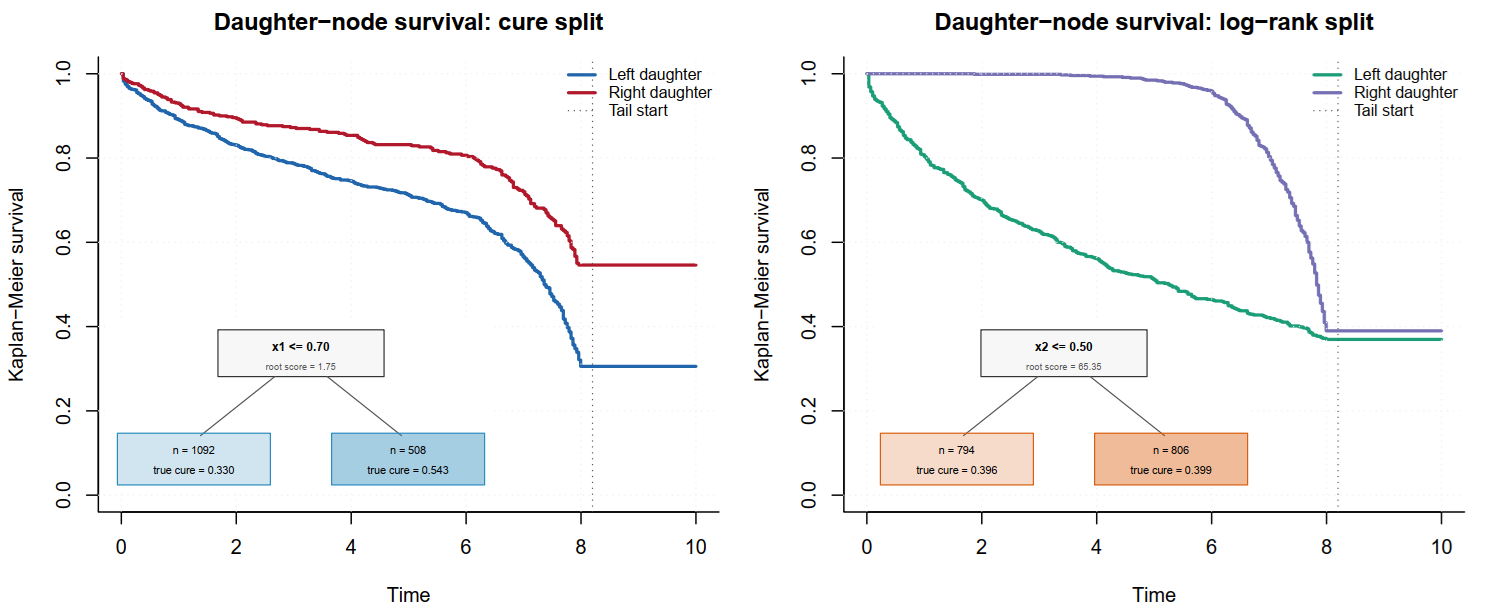}
\caption{ \textbf{Why a splitting criterion targeting cure fraction?}: The log-rank based criterion selects $X_2$, which affects only the timing of events among susceptible subjects, and yields daughters with different curves but nearly the same plateau. The cure-fraction based splitting criterion selects $X_1$, which affects the cure probability, and yields daughters with different plateaus.}
\label{fig:cure_vs_logrankprop}
\end{figure}

To illustrate why targeting the cure fraction matters.  we simulate data (n=1600 observations) in which \(X_1\) affects only the cure probability, whereas \(X_2\) affects only the survival times of susceptible subjects.Figure~\ref{fig:cure_vs_logrankprop} shows that the CBST selects the plateau variable \(X_1\), whereas the log-rank tree selects the variable \(X_2\), which drives the latency. The true cure contrast between daughter nodes is 0.214 for the cure-fraction based splitting criterion, but only 0.002 for the log-rank split: although the log-rank daughters have markedly different survival curves, their plateaus are nearly identical, so the split carries almost no information about cure. A criterion that targets the plateau directly avoids this problem.

\section{Cure-fraction based survival trees (CBST)}
\label{sec:trees}

 For a parent node \(A\), a binary split \(s=(j,c)\) partitions the covariate space into left and right daughter nodes, defined respectively as
\[
A_L(s)=\{x\in A:x_j\leq c\},\qquad
A_R(s)=\{x\in A:x_j>c\}\ .
\]
Unlike conventional survival trees based on traditional log-rank scores, our candidate partitions are evaluated using our proposed cure-fraction based splitting criterion. This choice ensures that the recursive partitioning process prioritizes the identification of strata with distinct probabilities of long-term survival. Heterogeneity in survival time of susceptible subject (latency) is subsequently estimated within the resulting terminal nodes once the long-term survivor structure has been isolated.

\noindent\textbf{Node-level cure estimation}:\label{sec:node_censoring}
 To evaluate candidate splits, we consider two node-level survival estimators depending on the underlying censoring mechanism. The default estimator is the ordinary Kaplan--Meier estimator, denoted by \(\widehat S_A^{\mathrm{KM}}\).
This estimator assumes nodewise independent censoring, that is \(T \perp C \mid X \in A\) for every admissible node \(A\). While this condition automatically holds when censoring is completely independent of both \(T\) and \(X\), it does not generally follow from the weaker global assumption \(T \perp C \mid X\).
When censoring depends on $X$, we replace the Kaplan–Meier curve with a cross-fitted IPCW node
estimator, which requires uniform consistency and positivity of the censoring model (explained later). %
Node consistency permits a nuisance fitted
outside the tree, but the forest Central Limit Theory (CLT) and ordinary Infinitesimal Jackknife (IJ) require it to be known,
external, or estimated wholly within each tree subsample.

For each node \(A\), we define a tail set \(\mathcal{T}_A\) containing
mature follow-up times. In the simplest situation, the tail estimate is
\[
\widehat\pi_A^{\mathrm{tail}}=\widehat S_A(\tau_A),
\]
where \(\tau_A\) is a pre-specified time-point. 
A more stable estimator averages the
node survival curve over a tail window:
\begin{equation}
\widehat\pi_A^{\mathrm{tail}}
=
\int_{\mathcal{T}_A}\widehat S_A(t)\,dW_A(t),
\label{eq:tail_pi_estimator}
\end{equation}
where \(W_A\) is a probability distribution supported on \(\mathcal{T}_A\).
In practice, \(W_A\) can be chosen to be uniform over a small grid of late survival times or over a fixed long-term window.

We adopt a tail average estimator to avoid large finite-sample 
sensitivity to isolated late events or censoring times. For a fixed window, however, the
target is a tail-survival functional and not necessarily the exact cure
fraction. Its difference from \(\pi_A\) is the residual susceptible survival
over that window, as shown in Theorem~\ref{thm:node_consistency}. The window may
be chosen using subject-matter knowledge or by a restricted data-adaptive rule.
Any adaptive rule used in the theory must satisfy the uniform conditions in
Assumption~\ref{ass:regularity}; a visually chosen plateau is not covered by
the formal results.

Tail support serves two distinct purposes. For the cure-fraction based splitting criterion, candidate
daughters are admissible only when they satisfy lower bounds on node size,
events before the tail, numbers at risk over \(\mathcal{T}_A\), and
censoring support. The log-rank comparator enforces the common daughter-size
constraint during candidate scoring but does not apply these cure-specific
event, late-risk-set, or censoring-support filters. Separately, a current
estimation node whose own tail is immature may inherit the cure prediction of
its nearest mature ancestor while structure growth continues if admissible
daughters exist. Such fallback is reported explicitly: it is not a node-
specific exact-cure estimate, and structure growth must not be conflated with
terminal-tail maturity.

Two projections have different purposes and are kept distinct. The
probability-valued terminal estimate is
\[
\overline\pi_A=
[\widehat\pi_A^{\mathrm{tail}}]_{[0,1]}\ ,
\]
where \([u]_{[a,b]}=\min\{b,\max(a,u)\}\). Only the split score uses the
interior clipping
$\widetilde\pi_A
=
[\overline\pi_A]_{[\varepsilon,1-\varepsilon]}$ for $
0<\varepsilon<1/2$.
The fixed interior clipping keeps the logit score finite; it is not used as the
terminal prediction because doing so would create a nonvanishing boundary bias
when the true cure probability is below \(\varepsilon\) or above
\(1-\varepsilon\).

\noindent\textbf{Cure-fraction based splitting criterion}: 
The following score has four components. The squared contrast in the numerator measures separation in long-term survival between two daughter nodes; the balance factor \(\omega_s=n_L n_R/n_A^2\) discourages very small daughters; the variances in the denominator account for the uncertainty in the tail estimates; and the stabilizer $\lambda>0$ prevents an unstable variance estimate from dominating the split.
The primary split score is
\begin{equation}
G_{\mathrm{cure}}(s;A)
=
\omega_s
\frac{
\left[
\logit\{\widetilde\pi_{A_L(s)}\}
-
\logit\{\widetilde\pi_{A_R(s)}\}
\right]^2
}{
\widehat V\{\logit(\widetilde\pi_{A_L(s)})\}
+
\widehat V\{\logit(\widetilde\pi_{A_R(s)})\}
+
\lambda
}\ ,
\label{eq:cure_split_score}
\end{equation}
where  
\(\widehat V\{\cdot\}\) is either a Greenwood or empirical variance
estimator.
For KM tails, the normalizer retains covariance across all grid times through
Greenwood increments; for IPCW tails, it is the empirical variance of the
weighted subject contributions divided by node size. The logit delta method is
then applied after clipping at fixed interior bounds. These choices stabilize
split ranking but remove neither boundary nonregularity nor uncertainty from an
estimated censoring model; formulas and the corresponding inferential
restriction are given in Supplementary Appendix~A.
The selected split is
$\widehat s_A=\arg\max_{s\in\mathcal{S}_A}G_{\mathrm{cure}}(s;A),
$
over the admissible candidate splits \(\mathcal{S}_A\). Unlike a log-rank
criterion, Equation~\eqref{eq:cure_split_score} is large when daughter nodes differ in
their long-term survivor probabilities.

The logit scale expresses separation on an odds scale and permits approximate
variance standardization through the Greenwood or empirical-contribution
normalizers above. It also
amplifies changes near 0 and 1, so the fixed clipping is a numerical
regularization rather than a remedy for boundary nonregularity. The split
recovery theory therefore assumes that the relevant population daughter-node
cure probabilities are uniformly separated from 0 and 1 and that
\(\varepsilon\) lies inside that separation. The balance factor discourages
splits that isolate a small number of subjects with unstable tails, and the
stabilizer \(\lambda\) prevents a nearly zero estimated variance from
dominating the ranking.

An integrated tail-distribution score and a cure--latency hybrid are optional
extensions when either a single plateau estimate is unstable or both components are
scientifically important. The theory and experiments here concern only
\(G_{\mathrm{cure}}\); the extensions require separate population targets and
uniform score laws.

Equation~\eqref{eq:cure_split_score} operationalizes the distinction illustrated
in Figure~\ref{fig:cure_vs_logrankprop}: the score rewards separation
in the estimated plateau, while its balance and variance terms discourage a
large contrast supported by only a small or unstable late risk set.

\noindent\textbf{Terminal-node estimation}:
For a subject assigned to terminal node \(\mathcal{A}(x)\), set
\(\widehat\pi_T(x)=\overline\pi_{\mathcal{A}(x)}\). When this quantity is below
one, reconstruct susceptible survival by
\begin{equation}
\widehat S_{u,\mathcal{A}(x)}(t)
=
\left[
\frac{\widehat S_{\mathcal{A}(x)}(t)
      -\overline\pi_{\mathcal{A}(x)}}
     {1-\overline\pi_{\mathcal{A}(x)}}
\right]_{[0,1]}.
\label{eq:node_latency_rescale}
\end{equation}
and recompose
\(\widehat S_T(t\mid x)=\widehat\pi_T(x)+
\{1-\widehat\pi_T(x)\}\widehat S_{u,T}(t\mid x)\).
If \(\widehat\pi_T(x)=1\), susceptible survival is not identified and is not
reported. This plug-in reconstruction is secondary to the cure estimate and
must be accompanied by terminal-tail support.

\section{Random forests of CBST}
\label{sec:bagging_rf}

Averaging perturbed trees stabilizes recursive partitioning
\citep{Breiman1996Bagging}; random candidate variables add the usual forest
randomization \citep{breiman2001}. Here every constituent tree is a CBST,
grown to separate probabilities of cure (long-term survival). The method is therefore not a standard
survival forest evaluated only with a long monitoring window: the estimand of long-term survival drives
the partitioning tool.

\noindent\textbf{Forest estimators}:
For tree-level cure estimates \(\widehat\pi_b(x)\), the CBST forest estimate is
\[
\widehat\pi_{\mathrm{CBST}}(x)
=
\frac{1}{B}\sum_{b=1}^{B}\widehat\pi_b(x).
\]
Population survival is defined by first constructing each tree's coherent
mixture curve and then averaging those curves. This order is essential:
projecting an average raw node-survival curve above the average cure estimate
is not generally equal to averaging coherent tree-specific mixture curves.
When the forest cure estimate is below one, subtracting it from the coherent
population-survival average and dividing by its complement yields a weighted
average of identifiable tree-level susceptible curves. The dynamic cure
probability is \(\widehat q(t,x)=\widehat\pi_{\mathrm{CBST}}(x)/
\widehat S_{\mathrm{CBST}}(t\mid x)\). Derivations and the required aggregation
order are in Supplementary Appendix~A.

\noindent\textbf{Tail-survival and cure probability}:\label{sec:asymptotic_framework}
We first state what the forest estimates from the available follow-up.
Let $W_n$ be a common probability measure supported on a prespecified tail
set $\mathcal T_n$. The tail-survival target is given by
\begin{equation}
\psi_n(x)=\int_{\mathcal T_n}S(t\mid x)\,dW_n(t)
=\pi(x)+\beta_n(x),
\qquad
\beta_n(x)=\{1-\pi(x)\}\int_{\mathcal T_n}S_u(t\mid x)\,dW_n(t)\ge0.
\label{eq:finite_tail_point_target}
\end{equation}
For a node $A$ with positive probability,
\[
\psi_{A,n}=\int_{\mathcal T_n}S_A(t)\,dW_n(t)
=E\{\psi_n(X)\mid X\in A\},
\qquad
\beta_{A,n}=\psi_{A,n}-\pi_A\ge0.
\]
The common tail rule makes the node target an average of the same function
$\psi_n(x)$. A node-specific adaptive window can be used for descriptive
fitting, but a pointwise consistency result then requires an additional
condition linking those windows to a common target.

The difference between tail-survival and cure probability is the contribution
of susceptible subjects who remain event-free over the selected window.
Consequently, accurate estimation of tail-survival does not by itself identify
cure probability. Theorem~\ref{thm:node_consistency} in Appendix~A separates
the node-estimation error from this residual-tail bias. Tail-support
constraints help provide reliable estimates, but do not ensure that the bias
vanishes. Under insufficient follow-up, identification of cure requires
support or extrapolation assumptions beyond consistency on a fixed interval
\citep{maller1996,AmicoVanKeilegom2018,EscobarBachVanKeilegom2019}.

The following proposition explains why the split target matters. It gives
the population counterpart of Figure~\ref{fig:cure_vs_logrankprop}: a covariate
can separate susceptible event times without separating cure probabilities.

\begin{proposition}[Cure and latency can select different roots]
\label{prop:logrank_latency}
Suppose $X_1\perp X_2$; $X_1$ changes only the cure probability through an
effect $\delta>0$; and $X_2$ changes only susceptible event timing through a
fixed effect $\beta\ne0$. Assume identifiable follow-up, a common censoring
regime across candidate daughters, continuity of both population log-rank
criteria in $\delta$, and a nondegenerate latency contrast satisfying
$\mathcal L_2(0,\beta)>0$. Then the exact-cure criterion prefers $X_1$, whereas
for each fixed $\beta\ne0$ there is a $\delta_0(\beta)>0$ such that the
population log-rank criterion prefers $X_2$ whenever
$0<\delta<\delta_0(\beta)$.
\end{proposition}

For a fixed latency effect, the two criteria can therefore select different
variables when the cure effect is sufficiently small. The result concerns
their population targets; it is not a claim that every finite-sample CBST
split recovers the cure variable.

\noindent\textbf{Recovery of the split}:
We now consider the finite set of cutpoints evaluated at a parent node $A$.
For a candidate split $s$, let $p_D(s)$ be the conditional population mass of
daughter $D\in\{L,R\}$ and let $\nu_{D,n}(s)$ be the population counterpart of
its variance normalizer. Define
$\ell_\varepsilon(u)=\logit([u]_{[\varepsilon,1-\varepsilon]})$. The population
score for tail-survival is
\begin{equation}
G^*_{\psi,n}(s;A)
=p_L(s)p_R(s)
\frac{\{\ell_\varepsilon(\psi_{L,n}(s))-
\ell_\varepsilon(\psi_{R,n}(s))\}^2}
{\nu_{L,n}(s)+\nu_{R,n}(s)+\lambda}.
\label{eq:population_finite_tail_score}
\end{equation}
The score $G^*_\pi(s;A)$ is obtained by replacing the two tail-survival targets
with the two cure probabilities.

For the following result, we use (A1)--(A4), the tail-survival regime in (B1),
and (C1)--(C3), stated in Appendix~A. These conditions specify response
honesty, uniform node estimation, daughter balance, an interior target,
and a positive score stabilizer. They also require a unique population
maximizer $s^*_{\psi,A,n}$ with score gap $\gamma_{A,n}$, and uniform estimation
error of order $\rho_{A,n}$ satisfying $\rho_{A,n}/\gamma_{A,n}\to0$.
Let $\widehat s_A$ maximize the empirical score over $\mathcal S_A$.

For recovery of the cure split, we additionally require that $G^*_\pi$ have a
unique maximizer $s^*_{\pi,A,n}$ on the same candidate set, with gap
$\gamma^\pi_{A,n}$, and that
\[
\rho_{A,n}+\max_{s,D}\beta_{D,n}(s)
=o(\gamma^\pi_{A,n}).
\]
We refer to these as the additional cure-recovery conditions.

\begin{theorem}[Recovery of the tail-survival and cure splits]
\label{thm:split_consistency}
Under the stated tail-survival conditions,
\[
\sup_{s\in\mathcal S_A}
|G_{\mathrm{cure}}(s;A)-G^*_{\psi,n}(s;A)|
=O_p(\rho_{A,n}),
\qquad
P\{\widehat s_A=s^*_{\psi,A,n}\}\longrightarrow1.
\]
Under the additional cure-recovery conditions,
$P\{\widehat s_A=s^*_{\pi,A,n}\}\to1$.
\end{theorem}

The obtained split therefore approaches the best split for the observable
tail-survival target. To recover the best cure split, both estimation error
and residual-tail bias must be small relative to the difference between
competing cure scores. Without the additional rate condition, recovery is
for the tail-survival score only.

Corollaries~\ref{cor:empirical_cutpoints} and~\ref{cor:null_stopping} in
Appendix~A extend this argument to empirical cutpoint grids and stopping rules.
A sufficiently rich grid can approximate the optimum of a continuous split
class. A suitable threshold can prevent splitting when the population score
is zero. The corresponding cure conclusions also require control of
residual-tail bias relative to the score gap or threshold. These rate
conditions do not provide a universal numerical threshold; the simulation
calibration concerns the finite-sample root decision only.

\noindent\textbf{Consistency of forest predictions}:
The prediction from a tree may come from its terminal leaf or from a mature
ancestor. Let $\mathcal P_{n,\Theta}(x)$ denote this effective prediction node
for a randomized tree $\Theta$, and let $\widehat\psi_{n,\Theta}(x)$ be its
returned estimate. Its conditional mean is
\[
m_n(x)=E_\Theta\{\widehat\psi_{n,\Theta}(x)\mid\mathcal D_n\}.
\]
The forest error has three components: Monte Carlo error from using finitely
many trees, estimation error within the effective nodes, and localization
error between those nodes and the point $x$. Denote them by
$M_{n,B_n}(x)$, $R_{n,\mathrm{est}}(x)$, and $R_{n,\mathrm{loc}}(x)$,
respectively. Their exact definitions and decomposition are given in
Equation~\eqref{eq:forest_consistency_decomp} in Appendix~A.

We use conditions (D1)--(D3): conditionally independent bounded trees,
consistent estimation within the effective nodes, and localization of their
tail-survival targets at $x$. Shrinking cells and local equicontinuity provide
a sufficient localization condition; persistent fallback to a coarse ancestor
can prevent it. These properties are assumptions of the result, not automatic
consequences of the splitting rule. For consistency of the forest average,
we also require $B_n\to\infty$. The cure conclusion additionally requires
$\beta_n(x)\to0$.

\begin{theorem}[Pointwise forest consistency]
\label{thm:forest_consistency}
Under (D1)--(D3), the error components satisfy
\[
M_{n,B_n}(x)=O_p(B_n^{-1/2}),\qquad
R_{n,\mathrm{est}}(x)=o_p(1),\qquad
R_{n,\mathrm{loc}}(x)=o_p(1).
\]
With the stated condition on $B_n$, the forest is consistent for tail-survival:
\[
\widehat\psi_{n,B_n}(x)-\psi_n(x)\xrightarrow{p}0.
\]
With the additional cure condition,
$\widehat\psi_{n,B_n}(x)\to_p\pi(x)$.
\end{theorem}

Increasing the number of trees reduces Monte Carlo error. It does not remove
node-estimation error, localization error, or residual-tail bias.
Theorem~\ref{thm:honest_tree_consistency} in Appendix~A gives the corresponding
fixed-depth tree result and distinguishes a terminal-node target from the
mature-ancestor target returned when fallback remains active.

Corollary~\ref{cor:derived_functionals} gives the conditions for consistency of
the population, susceptible, and dynamic survival predictions. It requires
reliable estimation and localization for the nodes supplying these quantities,
as well as cure consistency. Susceptible survival and dynamic cure probability
also require their denominators to be bounded away from zero. Without
vanishing residual-tail bias, substituting tail-survival for cure probability
in the mixture identities does not generally identify these derived quantities.

\noindent\textbf{Uncertainty in forest predictions}:\label{sec:forest_inference}
Inference requires more than consistency. Each tree must use one uniformly
drawn, fixed-size subsample without replacement, with $s_n\to\infty$ and
$s_n/n\to0$. Tail controls and censoring nuisances must be external or computed
within that subsample. A failed tree may return a prespecified bounded
fallback, but the construction does not allow rejection of the subsample and
repeated sampling until a tree succeeds. Full-data tail selection and a shared
estimated IPCW nuisance require an additional influence analysis.

Let $\theta_n(x)$ be the repeated-sample expectation of the subsampled forest
kernel and let $\sigma_n^2(x)$ be its first-order sampling variance.
Appendix~A defines these quantities and states conditions (I1)--(I4).
Conditions (I1)--(I3) give the sampling limit: they specify the subsampling
construction, dominance of the first Hoeffding projection with a Lindeberg
condition, and negligible finite-tree Monte Carlo error.
Condition (I4) additionally controls the inclusion-covariance approximation
used by the IJ. These conditions are not asserted for every adaptive tree
\citep{MentchHooker2016,WagerAthey2018}.

For $N_{bi}$ indicating inclusion of subject $i$ in tree $b$, define
\[
\widehat C_i(x)=B^{-1}\sum_{b=1}^B
\{N_{bi}-\overline N_i\}
\{\widehat\psi_b(x)-\overline T(x)\}.
\]
The fixed-size without-replacement IJ is
\[
\widehat V_{\mathrm{IJ}}(x)
=a_{n,s}\sum_{i=1}^n\widehat C_i^2(x),
\qquad
a_{n,s}=\frac{n-1}{n}\left(\frac{n}{n-s_n}\right)^2.
\]

For inference about tail-survival, the additional bias condition is
\[
\theta_n(x)-\psi_n(x)=o\{\sigma_n(x)\}.
\]
For inference about cure probability, we also require
\[
\beta_n(x)=\psi_n(x)-\pi(x)=o\{\sigma_n(x)\}.
\]
These conditions require the relevant biases to be negligible relative to the
standard error, rather than merely to converge to zero.

\begin{theorem}[Sampling limit and variance estimation]
\label{thm:inference_summary}
Under (I1)--(I3),
\[
\{\widehat\psi_{n,B_n}(x)-\theta_n(x)\}/\sigma_n(x)
\xrightarrow{d}N(0,1).
\]
Under the additional condition (I4),
\[
\widehat V_{\mathrm{IJ}}(x)/\sigma_n^2(x)\xrightarrow{p}1.
\]
Under the respective bias conditions stated above, the Gaussian limit also
holds when centered at $\psi_n(x)$ or at $\pi(x)$.
\end{theorem}

Theorem~\ref{thm:inference_summary} combines the sampling limit, target-bias,
and IJ results stated separately in Appendix~A. The IJ estimates variation
around the forest expectation; it does not estimate localization or residual-tail
bias. Thus, variance estimation alone does not justify an interval for cure
probability. Logit-scale inference further requires an interior target and
asymptotically inactive boundary projection. Paired contrasts use within-tree
prediction differences and their corresponding inclusion covariances.

For a finite number of trees, the software also reports the diagnostic
correction in Equation~\eqref{eq:finite_b_ij_correction}. This correction is
not implied by the IJ consistency result. It can exhaust the raw variance
when $B$ is too small and cannot remove localization, regularization,
nuisance-estimation, or residual-tail bias. For this reason, the simulations
evaluate variance calibration around the forest expectation separately from
coverage of the scientific target.

\section{Simulation study}
\label{sec:simulation}

\noindent\textbf{Design and evaluation}:\label{sec:simulation_objectives}
The simulation study evaluates recovery of cure heterogeneity when separate
covariates govern incidence and susceptible-event timing. Cure-probability
error and root-split recovery are the primary outcomes; Brier score and AUC
for latent cure status provide complementary measures. Additional designs on
depth, follow-up maturity, censoring, and inference are reported in the
Supplementary Material.

\label{sec:simulation_design}

We generated independent \(X_1\), \(X_2\), and noise \(X_3\), assigning
\(X_1\) to cure incidence and \(X_2\) to susceptible-event latency. Two
strengths for each signal, marginally independent and covariate-dependent
censoring, and \(n=500,1000,2000\)
gave 24 settings, with 100 paired repetitions and an independent test sample
of size 5000. KM and oracle-IPCW CBST forests were compared with a log-rank
forest using 300 honest trees, 80\% subsampling, maximum depth two, and a common
cutpoint grid, daughter-size controls, and tail interval. Candidates evaluated
with the cure-fraction based splitting criterion additionally satisfied tail-maturity filters. Complete distributions and
controls appear in Supplementary Appendix~A.

\noindent\textbf{Computational implementation}:
Structure outcomes determined splits, while a disjoint sample supplied
terminal estimates under prespecified feasibility and routing rules.
CBST fits used a positive score stabilizer and sampling without
replacement. Finite-time survival was made coherent within each tree before
averaging. All analyses used \texttt{cureforest} version 0.1.3, now distributed as the R package \texttt{curelearning}.

\label{sec:simulation_main_results}

The selected root variable reveals whether cure incidence or latency drives
the partition. Table~\ref{tab:main_confirmatory_results} averages equally over
signal strengths and censoring mechanisms, summarizing 800 paired data sets
per sample size.

\begin{table}[!htbp]
\centering
\scriptsize
\caption{Confirmatory cure--latency separation results. Entries are averages
over two cure strengths, two latency strengths, two censoring mechanisms, and
100 Monte Carlo repetitions per scenario. Signal, latency, and noise denote
the proportions of root splits on \(X_1\), \(X_2\), and \(X_3\), respectively.}
\label{tab:main_confirmatory_results}
\begin{tabular}{@{}rlrrrrrr@{}}
\toprule
\(n\) & Method & MSE\(_\pi\) & Brier & AUC & Signal & Latency & Noise \\
\midrule
500  & CBST KM       & 0.00606 & 0.21910 & 0.6612 & 0.899 & 0.029 & 0.073 \\
     & CBST IPCW     & 0.00917 & 0.22212 & 0.6507 & 0.756 & 0.069 & 0.175 \\
     & Log-rank KM   & 0.00990 & 0.22295 & 0.6553 & 0.707 & 0.260 & 0.033 \\
\addlinespace
1000 & CBST KM       & 0.00353 & 0.21665 & 0.6668 & 0.977 & 0.005 & 0.018 \\
     & CBST IPCW     & 0.00564 & 0.21876 & 0.6607 & 0.884 & 0.034 & 0.082 \\
     & Log-rank KM   & 0.00650 & 0.21962 & 0.6615 & 0.685 & 0.310 & 0.005 \\
\addlinespace
2000 & CBST KM       & 0.00247 & 0.21558 & 0.6679 & 0.997 & 0.001 & 0.002 \\
     & CBST IPCW     & 0.00355 & 0.21667 & 0.6644 & 0.964 & 0.011 & 0.026 \\
     & Log-rank KM   & 0.00511 & 0.21820 & 0.6640 & 0.678 & 0.322 & 0.000 \\
\bottomrule
\end{tabular}
\end{table}

prediction metrics. 
The distinction between the methods is more apparent in recovery of cure structure than in conventional predictive metrics. The probability that the KM CBST forest selected the true
cure variable increased from 0.899 at \(n=500\) to 0.997 at \(n=2000\). For the
log-rank forest, the corresponding probability remained between 0.678 and
0.707, while selection of the pure latency variable increased from 0.260 to
0.322. Thus, more data concentrated the proposed forest on the cure signal, but
did not remove the log-rank forest's attraction to event timing.

This structural difference translated into more accurate cure probabilities.
The KM CBST forest had the smallest MSE at every sample size and
reduced MSE relative to the log-rank comparator by 38.8\%, 45.7\%, and
51.7\% for \(n=500,1000,2000\), respectively. The Brier score also improved,
whereas the AUC differences were modest. This is the practical reason to report
split recovery together with prediction: two forests can rank subjects
similarly while forming different long-term survivor groups and producing different
cure-probability calibration.

The advantage was particularly clear under moderate cure heterogeneity
(\(a=1.5\)). Averaged over sample sizes, latency strengths, and censoring
mechanisms, cure-probability MSE was 0.00274 for the KM CBST forest and 0.00540
for the log-rank forest, while their cure-variable root-selection frequencies
were 0.920 and 0.456. Under the stronger cure signal (\(a=3.0\)), the
corresponding MSE values were 0.00530 and 0.00893, and the root-selection
frequencies were 0.995 and 0.924. The cure-fraction based splitting criterion therefore remained
most consequential when cure heterogeneity was present but not so strong that
both criteria selected the same variable.

Oracle IPCW selected the latency variable less often than log-rank splitting,
but had larger cure-probability MSE than KM in this design. This finite-sample
comparison does not establish a general preference against censoring
adjustment: KM requires plausible nodewise independent censoring, whereas IPCW
addresses dependence on observed covariates. The dedicated censoring
experiment in the Supplementary Material examines this distinction.

\noindent\textbf{Stopping under cure homogeneity}:
We next asked whether the cure-fraction based splitting criterion would create a split when the population
cure fraction was constant. For each sample size and censoring mechanism, the
threshold was fixed at the empirical 95th percentile of the root score of the cure-fraction based splitting criterion
from 500 null samples and then evaluated in 500 independently generated samples
at each cure-signal strength. Each fit was a single honest depth-one tree, so
this experiment evaluates root stopping rather than a complete node-adaptive
recursive procedure.

\begin{table}[!htbp]
\centering
\scriptsize
\caption{Independent validation of null-calibrated root stopping (500
calibration and 500 validation samples per row). Null split is the probability
of any root split when \(a=0\). The final columns give the probability of
recovering the cure variable \(X_1\) under weak, moderate, and strong cure
heterogeneity.}
\label{tab:null_stopping_results}
\begin{tabular}{@{}rlrrrrr@{}}
\toprule
\(n\) & Censoring & Threshold & Null split &
\(a=0.75\) & \(a=1.5\) & \(a=3\) \\
\midrule
500  & Covariate-dependent & 1.447 & 0.046 & 0.116 & 0.420 & 0.926 \\
     & Independent         & 1.380 & 0.066 & 0.154 & 0.424 & 0.954 \\
1000 & Covariate-dependent & 1.160 & 0.042 & 0.232 & 0.752 & 1.000 \\
     & Independent         & 1.256 & 0.040 & 0.196 & 0.710 & 1.000 \\
2000 & Covariate-dependent & 0.885 & 0.028 & 0.406 & 0.968 & 1.000 \\
     & Independent         & 0.906 & 0.054 & 0.374 & 0.962 & 1.000 \\
\bottomrule
\end{tabular}
\end{table}

The pooled null-split probability was \(138/3000=0.046\), ranging from 0.028
to 0.066 across cells; the pure-latency variable was selected in at most 0.006
of null samples. Moderate-signal recovery increased from 0.420--0.424 at \(n=500\) to
0.962--0.968 at \(n=2000\), and strong-signal recovery was at least 0.926 throughout.
Weak-signal recovery remained difficult, at 0.374--0.406 even at
\(n=2000\). Thus calibration
controlled spurious root splitting while retaining sensitivity to moderate
and strong cure heterogeneity. These thresholds are design-specific;
recursive use would require recalibration as node size and tail support
change.

\label{sec:simulation_inference}

The remaining designs and diagnostic tables are in Supplementary Appendix~A.
Allowing depth two reduced cure MSE from 0.00458 to 0.00346, with no material
gain at depth three.
In the nonlinear ten-covariate benchmark, the cure-fraction based splitting criterion reduced cure
MSE by 4.0--4.3\% and finite-time survival MSE by 0.9--3.5\% relative to the
log-rank comparator, with the latter advantage increasing with sample
size. A standard RSF had the lowest generic
prediction errors in that design, clarifying that the proposed method targets
tail-supported partitioning through CBST rather than universal predictive
dominance.

The inferential experiments used one-shot sampling without replacement,
sublinear subsample sizes, prespecified tails, and same-subsample fallback.
In the core profile, corrected IJ standard errors tracked variation around
the forest expectation (IJ/SD 0.971--1.059; centered coverage 0.920--0.960).
However, immature follow-up reduced exact-cure coverage despite root recovery
above 99.9\%. Independent confirmation showed decreasing localization bias
as \(n\) doubled, but the finite-\(B\) subtraction at \(n=2000\), \(B=1000\)
removed about 95\% of raw IJ variance and returned zero in about 5\% of
repetitions. These results distinguish variance calibration from target
coverage: bias must also be small relative to the standard error, and finite
forests must estimate inclusion covariances reliably. Under covariate-dependent
censoring, oracle IPCW changed the partitions without improving bias or RMSE
in the design considered. Variance calibration around the forest expectation
therefore remains distinct from routine exact-cure inference.

\section{FOSP data application}
\label{sec:fosp_application}

The proposed methodology is illustrated using the FOSP cancer registry. The
application asks a focused question: can the cure-fraction based splitting criterion form
long-term risk groups that remain separated in a later cohort while retaining
enough 8--10-year follow-up to estimate their tails directly? The analysis is
prognostic rather than causal, and full-curve prediction is secondary to this
tail-support question.

We used the public hospital cancer registry of the Funda\c{c}\~ao Oncocentro de
S\~ao Paulo (FOSP), downloaded May 27, 2026 \citep{FOSP2026,almeida2022fosp}.
Date, age, and follow-up checks retained 1,222,434 of 1,257,217 records; the
complete sequential flow appears in Supplementary Appendix~A.

We report two temporal case studies chosen for different scientific roles.
Breast cancer provides the largest eligible cohort and the strongest 10-year
risk-set support. Laryngeal cancer was selected in the 2008--2009 discovery
screen and carried forward without using its later outcomes. The results are
therefore temporally held out within each cancer site, although the displayed
sites should be interpreted as illustrations rather than a random sample of
cancer types.

Models were fitted to 2000--2009 diagnoses and evaluated without refitting in
2010--2013. Preprocessing and rare- or unseen-level mappings were learned from
training data only. Table~\ref{tab:fosp_cohorts} gives the analysis counts.

\begin{table}[!htbp]
\centering
\small
\caption{FOSP temporal case studies. Cancer deaths and numbers at risk refer to
the 2010--2013 temporal validation cohort.}
\label{tab:fosp_cohorts}
\begin{tabular}{lrrrr}
\toprule
Cancer group & Training & Validation & Cancer deaths & At risk at 10 years \\
\midrule
Breast (C50) & 52,604 & 32,645 & 7,802 & 10,439 \\
Larynx (C32) & 7,469 & 3,728 & 1,861 & 612 \\
\bottomrule
\end{tabular}
\end{table}

Time was measured from diagnosis. Cancer death was the primary event and
other-cause death was censored. The resulting product-limit tail concerns a
latent cancer-death time under the stated censoring assumptions, not
cumulative incidence or an observed biological-cure label. The tail window was
fixed at 8--10 years, and support was evaluated at both endpoints.

Predictors were age, sex, grouped stage, grade, basis of diagnosis, and
morphology group; treatment was excluded. Both forests used 1000 trees at
seeds 1101, 2202, and 3303, with honest 80\% subsamples and shared global
controls, including terminal estimation and fallback. During split selection,
only cure candidates additionally satisfied event and late-risk-set
constraints, so the comparison evaluates the complete partitioning rules.
An immature leaf used its nearest mature ancestor's tail estimate. Maximum
depth, cutpoints, and the remaining controls are specified in Supplementary
Appendix~A.

Figure~\ref{fig:fosp_breast_tree} first illustrates the two splitting targets
with a single breast-cancer tree at seed 1101. The CBST split
first on stage III and formed four terminal strata, whereas the log-rank tree
split first on stage IV and formed three strata. The terminal survival curves
were then estimated only in the 2010--2013 validation cohort. Across the three
prespecified seeds, the CBST had a mean Harrell C-index of 0.719,
compared with 0.650 for the log-rank tree, and a smaller 8-year IPCW Brier score,
0.145 compared with 0.165. The corresponding extreme-group tail contrasts were
0.460 and 0.297. Thus, the choice of splitting target produced clinically
interpretable long-term strata before forest aggregation was considered.

\begin{figure}[!htbp]
\centering
\includegraphics[width=\textwidth]{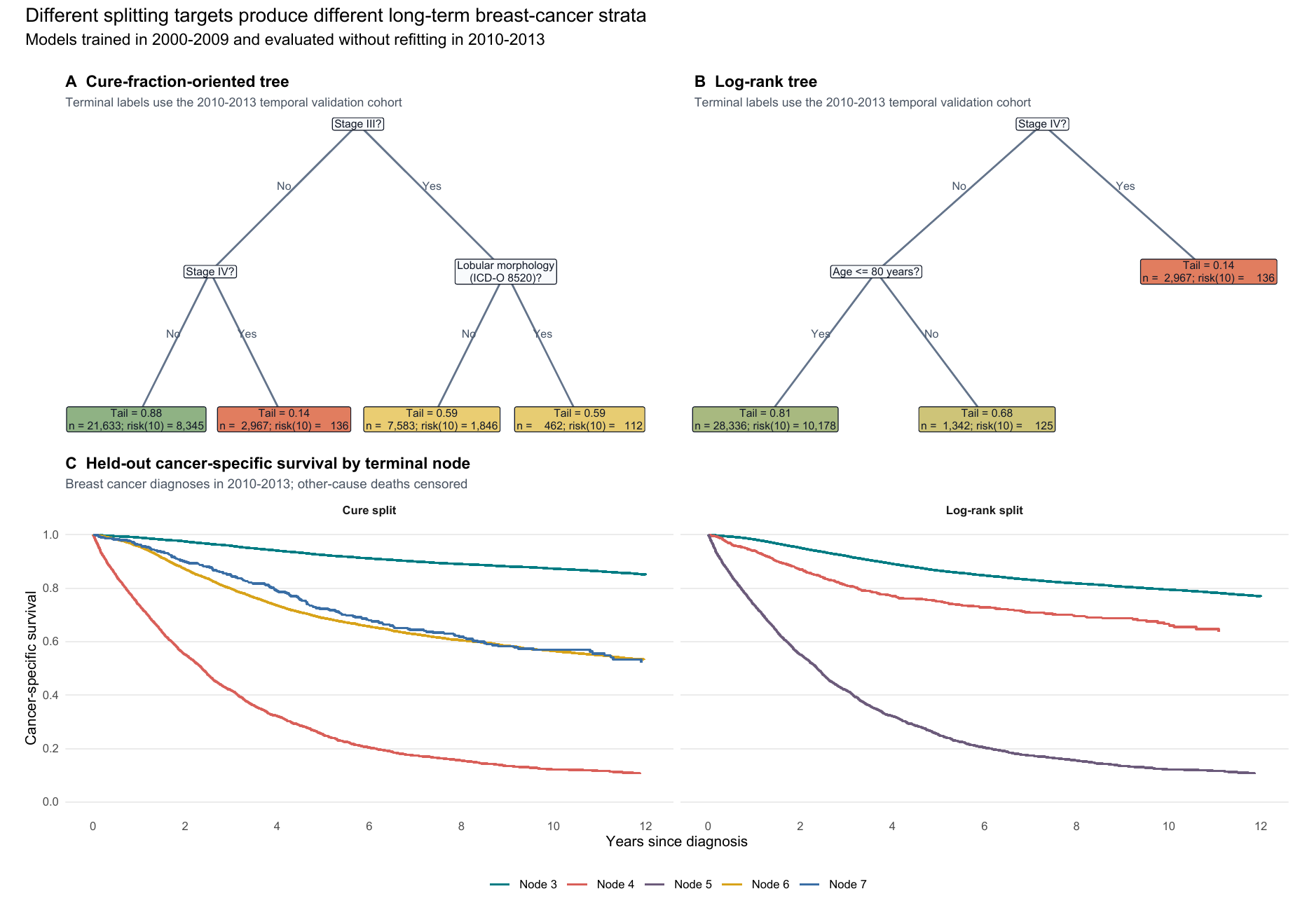}
\caption{Breast-cancer tree illustration. Panels A and B show the
CBST and log-rank trees fitted to 2000--2009 diagnoses at seed 1101;
terminal labels use only the 2010--2013 validation cohort. Panel C gives the
held-out cancer-specific survival curves for the corresponding terminal nodes.
Other-cause deaths were treated as censoring.}
\label{fig:fosp_breast_tree}
\par\small\textit{Alt text:} The CBST first separates stage III
and forms four strata of long-term survival, while the log-rank tree first
separates stage IV and forms three strata. Held-out survival curves show the
different long-term survival groups induced by the two splitting criteria.
\end{figure}

The primary summaries were held-out 8--10-year survival separation between
exactly disjoint extreme quintiles of predicted cure probability, their
numbers at risk at 10 years, and the fraction of terminal nodes requiring
fallback to a mature ancestor. Values below are averages over the three seeds.

\begin{figure}[!htbp]
\centering
\includegraphics[width=\textwidth]{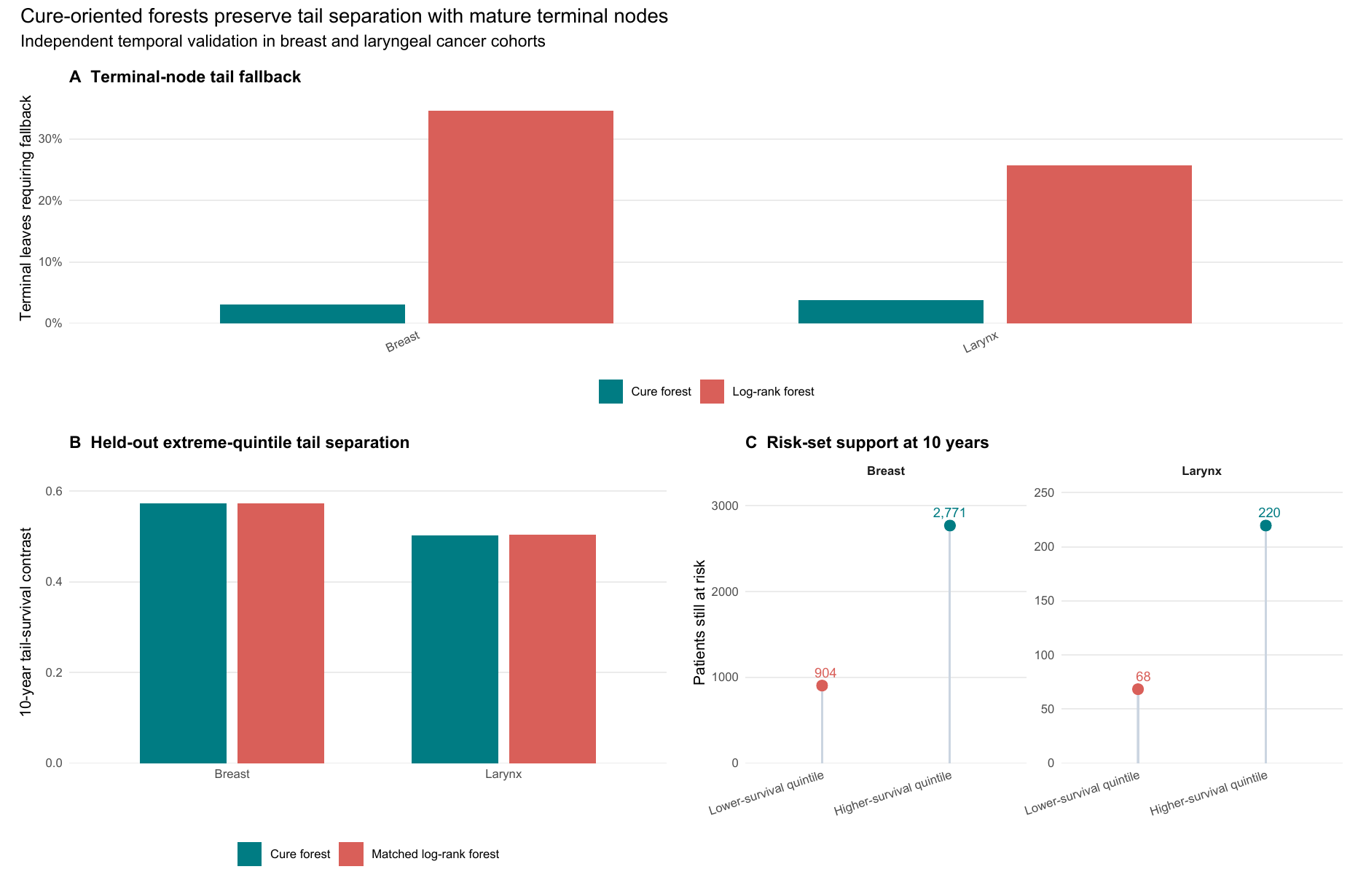}
\caption[Temporal validation of tail maturity and separation.]
{Temporal validation of tail maturity and separation. Panel A compares
terminal-node fallback, Panel B compares held-out extreme-quintile tail
separation, and Panel C reports 10-year risk-set support for the CBST forest.
Models were trained on 2000--2009 diagnoses and evaluated without refitting on
2010--2013 diagnoses. The figure label ``matched log-rank forest'' denotes the
shared global forest controls; candidate tail-maturity admissibility differs as
described in the text.}
\label{fig:fosp_validation}
\par\small\textit{Alt text:} Three-panel display for breast and laryngeal
cancer. CBST forests
have much lower fallback than log-rank comparators under shared global controls,
nearly identical
held-out tail-survival contrasts, and nonzero 10-year risk sets in both
extreme quintiles.
\end{figure}

In breast cancer, both forests produced a held-out tail contrast of 0.573.
Mature-ancestor fallback fell from 34.5\% under log-rank splitting to 3.1\%
under the cure-fraction based splitting criterion, a 91\% reduction. Thus, the CBST forest preserved the
separation while more often estimating the tail directly within its terminal
nodes.

The laryngeal cohort showed a similar pattern: tail contrasts were 0.503 and
0.504 for the CBST and log-rank forests, respectively, while fallback fell
from 25.7\% to 3.8\%, an 85\% reduction. Seed-to-seed standard deviations of
the CBST-forest tail contrast were 0.001 in breast cancer and 0.016 in
laryngeal cancer. Even in the smaller laryngeal cohort, the extreme
CBST-forest quintiles retained 68 and 220 subjects at risk at 10 years,
providing direct support for their tail-survival estimates.

Table~\ref{tab:fosp_validation} gives the corresponding 10-year risk sets and
terminal-node counts. Together with Figure~\ref{fig:fosp_validation}, it shows
that CBST forests used fewer terminal nodes while retaining
long-term separation and requiring less borrowing from mature ancestors.

\begin{table}[!htbp]
\centering
\small
\caption{Independent temporal validation, averaged over three 1000-tree
forests. Tail contrast is the difference in held-out 8--10-year survival
between disjoint extreme quintiles of predicted cure probability.}
\label{tab:fosp_validation}
\resizebox{\textwidth}{!}{%
\begin{tabular}{llrrrrr}
\toprule
Cancer & Split rule & Tail contrast & Risk set, lower & Risk set, higher &
Fallback (\%) & Mean terminal nodes \\
\midrule
Breast & CBST & 0.573 & 904 & 2,771 & 3.1 & 7.38 \\
Breast & Log-rank & 0.573 & 894 & 2,778 & 34.5 & 9.53 \\
Larynx & CBST & 0.503 & 68 & 220 & 3.8 & 5.80 \\
Larynx & Log-rank & 0.504 & 73 & 220 & 25.7 & 7.41 \\
\bottomrule
\end{tabular}}
\end{table}

The supplementary breadth analysis reports ten post-selected cancer sites
with C-index gains of 0.071--0.215 and 8-year Brier reductions of 3.7--25.6\%
relative to a single CBST. Selection used the same validation
outcomes, so these results are descriptive rather than independent confirmation.

Uncertainty intervals are not reported because these predictive fits used
80\% subsamples, whereas Theorems~\ref{thm:forest_clt}--\ref{thm:ij_consistency}
require a sublinear subsample sequence. Although IJ reproduced variability
around the forest expectation in the restricted simulation designs,
target-centered inference also requires negligible bias relative to the
standard error. The application fits therefore fall outside the interpretation
established by those results.

\section{Discussion}
\label{sec:conclusion}

In this paper, we proposed cure-fraction based survival trees (CBST) and their forests in which
contrasts in long-term survival guide recursive partitioning. The method
addresses studies where covariates may affect cure incidence and susceptible
event timing differently. By making tail-survival the splitting target,
the proposed forest forms groups intended to distinguish long-term survivor
probabilities.

The theoretical results explain when those probabilities can be interpreted
as cure. A tail estimator contains sampling error and a contribution from
susceptible subjects who remain event free over the chosen window. When this
residual component vanishes, exact-cure estimation becomes possible under
the stated split-recovery and localization conditions. Otherwise, the target
remains a tail-survival quantity. Follow-up maturity therefore determines
which interpretation is supported by the data. Effective-node localization
and reliable node estimation are also explicit conditions; they do not follow
automatically from increasing the number of trees.

The simulation study connects improved cure estimation to more frequent
recovery of the cure variable. The advantage over log-rank splitting was
largest when cure and latency had distinct covariates and smaller under
diffuse nonlinear effects. Modest AUC differences accompanied larger changes
in cure-probability error, showing why discrimination and structural recovery
should be assessed together. Independent null calibration also controlled
spurious root splitting while retaining sensitivity to moderate and strong
signals. This supports calibration at the root; recursive thresholds still
depend on node size and tail support.

The inference results distinguish estimation of variability from coverage of
the scientific target. IJ standard errors tracked sampling variation around
the forest expectation, but localization and residual-tail bias could still
produce substantial undercoverage. Regularization and estimation of a
censoring model can also contribute to target bias. Increasing the number of
trees reduces Monte Carlo error but does not by itself remove these sources
of error. Accurate split selection alone did not resolve that bias. The finite-\(B\) subtraction also became unstable in the
more aggressive subsampling configuration, so it remains an implementation
diagnostic. Reliable target inference requires both sufficiently accurate
variance estimation and bias that is small relative to the standard error.

The FOSP case studies illustrate the practical role of maturity constraints.
CBST forests retained long-term separation while reducing
mature-ancestor fallback by 85--91\%. More terminal regions could therefore
support their own tail estimates. The practical benefit is better use of late
follow-up within the learned groups, even when their overall separation is
similar. This gain concerns the complete partitioning rule, including its cure-specific support constraints. Interpretation remains
conditional on the hospital-registry population and the assumptions required
when other-cause deaths are censored; the resulting probability does not
identify individual biological cure.

The proposed methodology is most appropriate when long-term stratification is
the primary objective and mature follow-up is available. Standard survival
forests remain useful for overall event-time prediction, and hybrid criteria
may be appropriate when cure and latency are equally important. Extensions to
competing risks, relative survival, and bias-aware inference are priorities
for future work.

\section*{Software and reproducibility}

The implementation is now distributed as the R package \texttt{curelearning},
available at \url{https://github.com/ramospedroluiz/curelearning}.
The implementation includes tree-first coherent survival aggregation and the
restricted inference kernel. The reproducibility archive provides
replicate-level results, task manifests, diagnostics, and the manuscript
tables.

\section*{Disclosure statement}

The authors declare no conflicts of interest.

\section*{Data availability statement}

The de-identified FOSP Hospital Cancer Registry data are distributed
by the Funda\c{c}\~ao Oncocentro de S\~ao Paulo after registration on the
\href{https://fosp.saude.sp.gov.br/fosp/diretoria-adjunta-de-informacao-e-epidemiologia/rhc-registro-hospitalar-de-cancer/download-de-arquivos/}{FOSP registry download page} \citep{FOSP2026}.
The analysis used the snapshot downloaded on May 27, 2026. The raw registry
file is not redistributed with the article; reproduction of the application
requires registration, reacquisition of the registry data, and execution of
the documented preprocessing and analysis scripts with the version 0.1.3
implementation, now distributed as \texttt{curelearning}.

\appendix
\renewcommand{\thesection}{Appendix \Alph{section}}
\renewcommand{\thesubsection}{\Alph{section}.\arabic{subsection}}
\renewcommand{\thetheorem}{\Alph{section}.\arabic{theorem}}
\section{Mathematical conditions and proofs}
\label{app:proofs}

This appendix gives the conditions, supporting results, and proofs for
Section~\ref{sec:bagging_rf}. The forest consistency result assumes reliable
estimation and localization in the effective prediction nodes. It does not
establish these properties for every adaptive implementation. The inferential
results concern the one-shot subsampling construction specified in (I1).

\subsection{Conditions for consistency}
\label{app:consistency_conditions}

Let $\mathcal F_n^{\rm str}$ denote all structure-sample information and any
external controls used to construct a tree. Conditional on
$\mathcal F_n^{\rm str}$, let $\mathfrak A_n$ be the collection of nodes over
which a uniform statement is asserted, and set
\[
a_{A,n}=\sup_{t\in\mathcal T_n}
|\widehat S_A(t)-S_A(t)|.
\]

\begin{assumption}[Node, split, and forest conditions]
\label{ass:regularity}

We use four groups of conditions: node estimation, cure identification, split recovery, and forest localization:\\
\textbf{Group A:} (Sampling and censoring)
These conditions specify sampling, censoring, positivity, and node estimation.\\

\textnormal{(A1)} The observations are iid, and all tail, node, split,
fallback, and prediction maps are measurable.

\textnormal{(A2)} Trees are response-honest. Subjects used for terminal
estimation are disjoint from subjects whose outcomes determine cutpoints,
split scores, and stopping. Estimation covariates may enter only prespecified
feasibility and routing rules; statements involving those rules are conditional
on the covariate array.

\textnormal{(A3)} For the consistency results, $W_n$ is deterministic,
external, or selected using structure information only. For a fixed-tail
analysis, $\mathcal T_n\subset[0,\tau]$ for a deterministic $\tau<\infty$. A
moving-tail exact-cure analysis is allowed only when the strengthened uniform
survival-estimation condition in (A4) is verified on $\mathcal T_n$.

\textnormal{(A4)} For KM estimation, censoring is independent of failure within
each asserted node, node probabilities and at-risk probabilities are bounded
away from zero on the relevant time set, and honest estimation counts diverge.
For IPCW, $T\perp C\mid X$, the censoring survival is positive on the relevant
time set, fold-specific nuisance estimates are uniformly consistent, and the
weighted indicator class satisfies a uniform law of large numbers.
For a growing random node class or a moving tail, we additionally require
\[
\max_{A\in\mathfrak A_n}a_{A,n}=o_p(1).
\]

\textbf{Group B:} (Cure identification)
This condition specifies when the residual-tail bias must vanish.\\

\textnormal{(B1)} The tail-survival regime targets $\psi_{A,n}$ and needs no
claim that $\beta_{A,n}$ vanishes. Exact-cure statements require the relevant
tail bias to vanish: for split recovery,
$\max_{A\in\mathfrak A_n}\beta_{A,n}=o(1)$ at a rate stated below; for a
pointwise forest result, $\beta_n(x)\to0$. This can follow from a finite
susceptible endpoint lying inside the censoring support or from a moving tail
with sufficient at-risk support. A fixed bounded window does not generally
identify exact cure.

\textbf{Group C:} (Tree construction)
These conditions concern daughter-node mass, score estimation, and separation of the best split.\\

\textnormal{(C1)} Every daughter considered in a split has conditional mass at
least $c_0>0$. The population tail-survival targets used in the score lie in
$[\delta,1-\delta]$, where $0<\varepsilon<\delta<1/2$, and the score stabilizer
satisfies $\lambda\ge\lambda_0>0$.

\textnormal{(C2)} For each parent $A$, there is a deterministic sequence
$\rho_{A,n}\downarrow0$ such that, uniformly over its candidate daughters,
tail-survival estimates, daughter masses, and variance normalizers differ from
their population counterparts by $O_p(\rho_{A,n})$.

\textnormal{(C3)} Conditional on the structure information, the candidate
split set $\mathcal S_A$ used by the finite-sample recursion is finite. Its
tail-survival population score has a unique maximizer $s^*_{\psi,A,n}$ separated
from every other candidate by a gap $\gamma_{A,n}>0$ satisfying
$\rho_{A,n}/\gamma_{A,n}\to0$. A continuum population split class is handled
separately by Corollary~\ref{cor:empirical_cutpoints}, using neighborhood-wise
argmax separation rather than an impossible positive gap from arbitrarily close
cutpoints.

\textnormal{(C4)} Fixed-depth tree results involve only finitely many reached
nodes, and the preceding conditions hold at every population node in the
greedy recursion.

\textbf{Group D:} (Forest Localization)
These conditions concern estimation and localization in the nodes that supply the predictions.\\

\textnormal{(D1)} Conditional on the observed data and fixed controls, tree
randomizations are independent and identically distributed. Tree predictions
are bounded in $[0,1]$.

\textnormal{(D2)} For a fixed $x$, let
$\mathcal P_{n,\Theta}(x)$ be the node whose estimate is actually returned by
one randomized tree. It is the terminal cell when that cell is mature and the
nearest mature ancestor when fallback is used. Writing
$\widehat\psi_{n,\Theta}(x)$ for the returned estimate, assume
\[
E_\Theta\!
\left[
\left|\widehat\psi_{n,\Theta}(x)-
\psi_{\mathcal P_{n,\Theta}(x),n}\right|
\,
\middle|\,
\mathcal D_n
\right]=o_p(1).
\]
This is the random-effective-node estimation condition. It includes adequate
honest node size and tail support and does not silently treat an ancestor
fallback as a terminal-node estimate.

\textnormal{(D3)} The effective prediction nodes localize the tail-survival
target at $x$:
\[
E_\Theta\!
\left[
\sup_{z\in\mathcal P_{n,\Theta}(x)}
|\psi_n(z)-\psi_n(x)|
\,
\middle|\,
\mathcal D_n
\right]=o_p(1).
\]
Shrinking cells together with local equicontinuity of $\psi_n$ is a sufficient
condition. Persistent fallback to a coarse ancestor violates this condition.
\end{assumption}

For a fixed finite collection of nodes on a compact time set, the primitive
conditions in (A4) imply uniform node-survival consistency, as stated in
Lemma~\ref{lem:km_node_consistency}. For a growing random node class or a
moving tail, uniform convergence remains an additional requirement; it does
not follow from fixed-compact Kaplan--Meier theory alone.

\subsection{Node estimation and split recovery}
\label{app:node_split_results}

\begin{lemma}[Node-survival consistency]
\label{lem:km_node_consistency}
Under (A1)--(A4), for every fixed finite collection of response-honest nodes on
a compact time set,
\[
\max_{A\in\mathfrak A_n}a_{A,n}=o_p(1).
\]
The same conclusion on a growing random node class or a moving tail holds only
under the strengthened uniform condition explicitly required in (A4).
\end{lemma}

\begin{proof}
Fix a response-honest node $A$ and condition on the structure information that
defines it. For KM estimation on a compact time set, nodewise independent
censoring, positive node mass, positive at-risk probability, and diverging
honest estimation count imply uniform product-limit consistency,
\[
\sup_{t\in\mathcal T_n}
|\widehat S_A^{\rm KM}(t)-S_A(t)|=o_p(1).
\]
For a fixed finite node collection, taking the maximum preserves the conclusion.

For IPCW, condition fold by fold on the observations used to estimate the
censoring nuisance. Positivity and uniform nuisance consistency make replacement
of the estimated weight by the true weight uniformly negligible. With the true
weight,
\begin{align*}
E\left\{
\frac{\I(Y>t)}{G_C(t\mid X)}\mathrel{\Big|}X
\right\}
&=\frac{P(T>t,C>t\mid X)}{G_C(t\mid X)}\\
&=P(T>t\mid X)=S(t\mid X),
\end{align*}
where the second equality uses $T\perp C\mid X$. The weighted uniform law of
large numbers in (A4), followed by averaging over $X\mid X\in A$, gives the
same uniform conclusion. A finite maximum again proves the lemma.

Neither argument alone proves uniformity over an unrestricted growing random
node class or at moving endpoints. Those extensions are therefore required
explicitly in (A4).
\end{proof}

The following result separates node-estimation error from residual-tail bias.
It also shows that projection to the probability interval does not increase
the error for either target.

\begin{theorem}[Tail-survival target and residual-tail bias]
\label{thm:node_consistency}
For the unprojected node estimator
$\widehat\psi_{A,n}=\int\widehat S_A(t)\,dW_n(t)$,
\begin{equation}
\widehat\psi_{A,n}-\pi_A
=E_{A,n}+\beta_{A,n},
\qquad
|E_{A,n}|\le a_{A,n},
\label{eq:node_tail_decomposition_corrected}
\end{equation}
where
\[
\beta_{A,n}
=\{1-\pi_A\}\int S_{u,A}(t)\,dW_n(t)\ge0.
\]
Projection onto $[0,1]$ cannot increase absolute error relative to either
$\psi_{A,n}\in[0,1]$ or $\pi_A\in[0,1]$. Consequently,
$[\widehat\psi_{A,n}]_{[0,1]}-\psi_{A,n}=o_p(1)$ whenever
$a_{A,n}=o_p(1)$, and it is consistent for $\pi_A$ only when additionally
$\beta_{A,n}=o(1)$.
\end{theorem}

\begin{proof}
Add and subtract $S_A(t)$ inside the tail integral:
\begin{align*}
\widehat\psi_{A,n}-\pi_A
&=\int\{\widehat S_A(t)-S_A(t)\}\,dW_n(t)
  +\int\{S_A(t)-\pi_A\}\,dW_n(t)\\
&=E_{A,n}+\beta_{A,n}.
\end{align*}
Because $W_n$ is a probability measure, $|E_{A,n}|\le a_{A,n}$. The mixture
identity gives
\[
S_A(t)-\pi_A=\{1-\pi_A\}S_{u,A}(t)\ge0,
\]
which proves the formula and nonnegativity of $\beta_{A,n}$. Projection onto a
closed interval containing a target is nonexpansive. Thus convergence to
$\psi_{A,n}$ follows from $a_{A,n}=o_p(1)$, while convergence to $\pi_A$
requires the additional tail-bias condition.
\end{proof}

\noindent\textbf{Proof of Proposition~\ref{prop:logrank_latency}}:
\begin{proof}
For the split on $X_2$, independence of $X_1$ and $X_2$ implies equal marginal
cure probabilities in its two daughters, so exact-cure separation is zero. The
split on $X_1$ has daughter cure probabilities differing by $\delta$, so its
exact-cure separation is positive.

Let $\mathcal L_j(\delta,\beta)$ be the population squared log-rank criterion
for splitting on $X_j$ under the common censoring regime. At $\delta=0$, the
$X_1$ daughters have identical mixture survival distributions, so
$\mathcal L_1(0,\beta)=0$. The nondegenerate latency condition gives
$\mathcal L_2(0,\beta)>0$. Continuity in $\delta$ implies
$\mathcal L_1(\delta,\beta)\to0$ and
$\mathcal L_2(\delta,\beta)\to\mathcal L_2(0,\beta)>0$, yielding the stated
$\delta_0(\beta)$.
\end{proof}

\noindent\textbf{Proof of Theorem~\ref{thm:split_consistency}}:
\begin{proof}
On the interior interval $[\delta,1-\delta]$, the clipped logit map is uniformly
Lipschitz. Daughter masses and variance normalizers remain in compact sets, and
the score denominator is at least $\lambda_0$. Hence the map from the vector of
two daughter targets, two masses, and two normalizers to the split score is
uniformly Lipschitz. Assumption (C2) therefore gives
\[
\sup_{s\in\mathcal S_A}
|G_{\mathrm{cure}}(s;A)-G^*_{\psi,n}(s;A)|
=O_p(\rho_{A,n}).
\]
On the event that this supremum is smaller than half the tail-survival population
gap, no candidate other than $s^*_{\psi,A,n}$ can maximize the empirical
score. The probability of that event tends to one by (C3).

The same Lipschitz argument gives
\[
\sup_s|G^*_{\psi,n}(s;A)-G^*_{\pi}(s;A)|
=O\!\left(\max_{s,D}\beta_{D,n}(s)\right).
\]
If the sum of this deterministic target discrepancy and the empirical score
error is $o(\gamma^{\pi}_{A,n})$, the empirical score lies within half the
exact-cure gap of $G^*_{\pi}$ with probability tending to one. The exact-cure
maximizer is then selected.
\end{proof}

The finite candidate-set result extends to a continuous split class when the
empirical grid approximates the population optimum. The required separation
is stated for points outside each neighborhood of that optimum.

\begin{corollary}[Continuum split class and empirical cutpoints]
\label{cor:empirical_cutpoints}
Let $(\mathbb S_A,d)$ be a compact population split class and let
$s^*_{\psi,A,n}$ be its unique tail-survival maximizer. Suppose the empirical
score converges uniformly on $\mathbb S_A$, daughter balance is preserved,
and, for every $\epsilon>0$,
\[
\gamma_{A,n}(\epsilon)
=G^*_{\psi,n}(s^*_{\psi,A,n};A)
-\sup_{d(s,s^*_{\psi,A,n})\ge\epsilon}
 G^*_{\psi,n}(s;A)>0,
\]
with the uniform score error $o_p\{\gamma_{A,n}(\epsilon)\}$. Suppose also
that the empirical cutpoint grid contains a candidate whose population score is
within $o\{\gamma_{A,n}(\epsilon)\}$ of the optimum. Then every empirical
maximizer $\widehat s_A$ satisfies
$d(\widehat s_A,s^*_{\psi,A,n})\to_p0$. The analogous exact-cure conclusion
holds when the residual-tail score discrepancy is also
$o\{\gamma^{\pi}_{A,n}(\epsilon)\}$ for every $\epsilon>0$.
\end{corollary}

\begin{proof}
Fix $\epsilon>0$. By assumption, the population optimum exceeds the supremum
outside its $\epsilon$-neighborhood by $\gamma_{A,n}(\epsilon)$. The grid
contains a candidate whose population score is within
$o\{\gamma_{A,n}(\epsilon)\}$ of the optimum, and the empirical score error
is uniformly $o_p\{\gamma_{A,n}(\epsilon)\}$. Therefore, with probability
tending to one, that near-optimal grid candidate has a larger empirical score
than every grid point outside the $\epsilon$-neighborhood. Every empirical
maximizer is consequently inside the neighborhood. Since this holds for each
fixed $\epsilon>0$, the asserted argmax convergence follows. Adding the
uniform tail-survival--to--exact-cure score discrepancy gives the exact-cure
version.
\end{proof}

We next consider a node whose population split score is zero. The stopping
threshold must dominate the estimation error. For a cure interpretation,
it must also dominate the discrepancy between the two population targets.

\begin{corollary}[Stopping under a population-score null]
\label{cor:null_stopping}
Suppose $\sup_s G^*_{\psi,n}(s;A)=0$ and a split is made only when the empirical
maximum exceeds $\kappa_{A,n}$. If $\kappa_{A,n}\to0$ and
$\rho_{A,n}/\kappa_{A,n}\to0$, the node remains unsplit with probability
tending to one. If the tail-survival population maximum is bounded below by a
positive constant, it splits with probability tending to one. Under an
exact-cure population-score null, the corresponding no-split conclusion
requires
\[
\frac{\rho_{A,n}+\sup_s|G^*_{\psi,n}(s;A)-G^*_{\pi}(s;A)|}
{\kappa_{A,n}}\longrightarrow0,
\]
for which
\(\{\rho_{A,n}+\max_{s,D}\beta_{D,n}(s)\}/\kappa_{A,n}
\to0\) is sufficient. Under an exact-cure alternative whose population
maximum is bounded below by a positive constant, splitting occurs when the
tail-survival--to--exact-cure score discrepancy is \(o(1)\). A random
permutation threshold requires the corresponding conditional rate conditions.
\end{corollary}

\begin{proof}
Under the tail-survival population-score null, uniform convergence gives
$\widehat G_A^{\max}=O_p(\rho_{A,n})$. Since
$\rho_{A,n}/\kappa_{A,n}\to0$, the probability that the empirical maximum
exceeds the threshold tends to zero. Under a fixed positive tail-survival margin,
$\widehat G_A^{\max}$ converges to a positive value while
$\kappa_{A,n}\to0$, so splitting occurs with probability tending to one.
Under an exact-cure population-score null, the empirical maximum is bounded by
the tail-survival empirical score error plus
\(\sup_s|G^*_{\psi,n}(s;A)-G^*_{\pi}(s;A)|\); the displayed
\(o(\kappa_{A,n})\) condition therefore makes the split probability vanish.
Under a positive exact-cure alternative, an \(o(1)\) score discrepancy and
\(\kappa_{A,n}\to0\) imply splitting. The same comparison works for a random
threshold under the stated conditional rate conditions.
\end{proof}

\subsection{Trees, forests, and derived predictions}
\label{app:forest_results}

The fixed-depth result applies split recovery at the finitely many nodes
reached by the population recursion. The prediction is evaluated at the node
that supplies the estimate, including an ancestor when fallback is used.

\begin{theorem}[Fixed-depth response-honest trees]
\label{thm:honest_tree_consistency}
Under (A1)--(A4), (B1), and (C1)--(C4), the empirical fixed-depth partition converges in probability
to the population greedy partition defined by $G^*_{\psi,n}$. If the
exact-cure gap and tail-bias condition of
Theorem~\ref{thm:split_consistency} holds at every reached node, the same
partition converges to the exact-cure greedy partition. For a prediction point
$x$, let $\mathcal A_n(x)$ be its terminal cell and $\mathcal P_n(x)$ the
actual mature node supplying its estimate. Then
\[
\widehat\psi_T(x)-\psi_{\mathcal P_n(x),n}=o_p(1).
\]
If $P\{\mathcal P_n(x)=\mathcal A_n(x)\}\to1$, fallback is asymptotically
inactive and the limit is the terminal-node target. Otherwise the correct limit
is the mature-ancestor target, not the terminal-node target.
\end{theorem}

\begin{proof}
Proceed by induction over the fixed depth. At the root,
Theorem~\ref{thm:split_consistency} selects the population tail-survival split
with probability tending to one. Conditional on all earlier splits being
correct, each next empirical parent equals its population counterpart, so the
same argument applies. Only finitely many nodes occur, and a union bound proves
convergence of the entire partition. Applying the second part of
Theorem~\ref{thm:split_consistency} at every node yields the exact-cure
partition statement.

Response honesty ensures that terminal estimation outcomes did not determine
the partition. Conditional on the partition and the covariate-only feasibility
rules, Theorem~\ref{thm:node_consistency} applies to every mature node in the
finite tree. The value returned at $x$ is computed in
$\mathcal P_n(x)$, whether that node is terminal or an ancestor. Therefore
$\widehat\psi_T(x)-\psi_{\mathcal P_n(x),n}=o_p(1)$. If fallback vanishes in
probability, $\mathcal P_n(x)=\mathcal A_n(x)$ with probability tending to one;
otherwise no terminal-node conclusion follows.
\end{proof}

For one randomized tree $\Theta$, let $\mathcal P_{n,\Theta}(x)$ be the
effective prediction node and let $\widehat\psi_{n,\Theta}(x)$ be its returned
estimate. With
$m_n(x)=E_\Theta\{\widehat\psi_{n,\Theta}(x)\mid\mathcal D_n\}$,
the error of an average of $B_n$ conditionally iid trees can be written as
\begin{align}
\widehat\psi_{n,B_n}(x)-\psi_n(x)
={}&\underbrace{\widehat\psi_{n,B_n}(x)-m_n(x)}_{M_{n,B_n}(x)}+\underbrace{E_\Theta[\widehat\psi_{n,\Theta}(x)-
\psi_{\mathcal P_{n,\Theta}(x),n}\mid\mathcal D_n]}
_{R_{n,\mathrm{est}}(x)}\\
&+\underbrace{E_\Theta[\psi_{\mathcal P_{n,\Theta}(x),n}-
\psi_n(x)\mid\mathcal D_n]}_{R_{n,\mathrm{loc}}(x)}.
\label{eq:forest_consistency_decomp}
\end{align}
This is an exact identity. Fallback is represented through the effective node,
rather than treated as a terminal-node estimate.

\noindent\textbf{Proof of Theorem~\ref{thm:forest_consistency}}:
\begin{proof}
Equation~\eqref{eq:forest_consistency_decomp} follows by adding and subtracting
$m_n(x)$ and the population tail-survival target of the random effective node.
Conditional on $\mathcal D_n$, the bounded tree predictions are iid, so
\[
\operatorname{Var}\{M_{n,B_n}(x)\mid\mathcal D_n\}\le\frac{1}{4B_n},
\]
and $M_{n,B_n}(x)=O_p(B_n^{-1/2})$.

Assumption (D2) and Jensen's inequality give
$|R_{n,\mathrm{est}}(x)|=o_p(1)$. Because
\[
|\psi_{\mathcal P,n}-\psi_n(x)|
\le
\sup_{z\in\mathcal P}|\psi_n(z)-\psi_n(x)|,
\]
(D3) gives $R_{n,\mathrm{loc}}(x)=o_p(1)$. Combining the three terms proves
tail-survival consistency. Finally,
\[
\widehat\psi_{n,B_n}(x)-\pi(x)
=\{\widehat\psi_{n,B_n}(x)-\psi_n(x)\}+\beta_n(x),
\]
so $\beta_n(x)\to0$ yields exact-cure consistency.
\end{proof}

The next result concerns the secondary predictions obtained from the mixture
identity. It requires estimation and localization for both the raw survival
nodes and the effective nodes supplying the tail estimates.

\begin{corollary}[Population, susceptible, and dynamic survival]
\label{cor:derived_functionals}
Let $\mathcal G$ be a fixed finite time grid. For one tree, let
$\mathcal A_{n,\Theta}(x)$ be the terminal cell supplying its raw survival
curve and let $\mathcal P_{n,\Theta}(x)$ be the mature node supplying its
tail-survival estimate. Define the tree-coherent curve by
\[
\widehat S^{\rm coh}_{n,\Theta}(t\mid x)
=\max\{\widehat S_{\mathcal A_{n,\Theta}(x)}(t),
          \widehat\psi_{n,\Theta}(x)\},
\qquad t\in\mathcal G,
\]
and average these coherent curves across trees. Suppose the exact-cure
conclusion of Theorem~\ref{thm:forest_consistency} holds and
\[
E_\Theta\!\left[
\max_{t\in\mathcal G}
|\widehat S_{\mathcal A_{n,\Theta}(x)}(t)-
  S_{\mathcal A_{n,\Theta}(x)}(t)|
+|\widehat\psi_{n,\Theta}(x)-
  \psi_{\mathcal P_{n,\Theta}(x),n}|
\,\middle|\,\mathcal D_n\right]=o_p(1),
\]
\[
E_\Theta\!\left[
\max_{t\in\mathcal G}
\sup_{z\in\mathcal A_{n,\Theta}(x)}|S(t\mid z)-S(t\mid x)|
+\sup_{z\in\mathcal P_{n,\Theta}(x)}|\psi_n(z)-\psi_n(x)|
\,\middle|\,\mathcal D_n\right]=o_p(1).
\]
Then the average of the tree-coherent curves is jointly consistent for
$\{S(t\mid x):t\in\mathcal G\}$. The same pointwise limit also holds for
the post-aggregation projection
$\max\{B_n^{-1}\sum_b\widehat S_b(t\mid x),
B_n^{-1}\sum_b\widehat\psi_b(x)\}$ under these assumptions, although the
two estimators are not equal at finite $B_n$. Where $1-\pi(x)$ and
$S(t\mid x)$ are bounded away from zero, the susceptible survival and dynamic
cure probability obtained from the mixture identities are jointly consistent
on $\mathcal G$. If $\beta_n(x)$ does not vanish, plugging the tail-survival
quantity $\psi_n(x)$ into those identities does not generally identify
$S_u(t\mid x)$ or $q(t,x)$.
\end{corollary}

\begin{proof}
The map $(u,v)\mapsto\max(u,v)$ is one-Lipschitz under the sup norm. The first
display therefore replaces the estimated terminal survival and effective-node
tail estimate by their population node counterparts with average error
$o_p(1)$. The second display localizes those two population quantities at
$S(t\mid x)$ and $\psi_n(x)$, uniformly over the finite grid. Conditional
Monte Carlo averaging contributes $O_p(B_n^{-1/2})$. Thus the coherent forest
curve converges jointly on $\mathcal G$ to
$\max\{S(t\mid x),\psi_n(x)\}$. Exact-cure consistency gives
$\psi_n(x)\to\pi(x)$, and the mixture model implies
$S(t\mid x)\ge\pi(x)$, so this limit is $S(t\mid x)$. The same Lipschitz
argument applied after averaging proves consistency of the post-aggregation
projection, without claiming finite-sample equality of the two aggregation
orders.

The exact-cure conclusion also supplies
$\widehat\pi(x)\to_p\pi(x)$. The mappings
\[
(\pi,S)\mapsto\frac{S-\pi}{1-\pi},
\qquad
(\pi,S)\mapsto\frac{\pi}{S}
\]
are continuous where their denominators are bounded away from zero, so the
continuous mapping theorem gives the susceptible-survival and dynamic-cure
claims. If $\widehat\pi$ converges instead to $\psi_n\ne\pi$, these mappings
converge to pseudo-functionals and not generally to $S_u$ or $q$.
\end{proof}

\subsection{Conditions for inference}
\label{app:inference_conditions}

A tree used for inference must be a kernel of one uniformly drawn fixed-size
subsample. All tail controls and censoring nuisances must be external or
functions of that same subsample. A failed tree may return a prespecified
bounded fallback for the same subsample. Drawing a new subsample until a tree
succeeds changes the inclusion law and is not covered by these results.

Write the prediction from a size-$s_n$ subsample as
\[
T_{n,x}(O_{i_1},\ldots,O_{i_{s_n}};\Xi),
\]
where $\Xi$ is independent internal randomization, and define the symmetric
kernel
\[
h_{n,x}(o_1,\ldots,o_{s_n})
=E_\Xi\{T_{n,x}(o_1,\ldots,o_{s_n};\Xi)\}.
\]
Let
\[
\theta_n(x)=E\{h_{n,x}(O_1,\ldots,O_{s_n})\},\qquad
 g_{n,x}(O_1)=E\{h_{n,x}\mid O_1\}-\theta_n(x),
\]
\[
\zeta_{1,n}(x)=\operatorname{Var}\{g_{n,x}(O_1)\},\qquad
\sigma_n^2(x)=\frac{s_n^2}{n}\zeta_{1,n}(x).
\]

\begin{assumption}[Additional inferential conditions]
\label{ass:forest_inference}
\textnormal{(I1)} Each tree uses exactly one uniformly drawn size-$s_n$
subsample without replacement, no rejection or retry resampling, and independent
internal randomization. The kernel is symmetric and bounded,
$s_n\to\infty$, and $s_n/n\to0$. Full-data tail selection, a shared estimated
IPCW nuisance, or any other full-data control is excluded unless its influence
is added separately.

\textnormal{(I2)} If $U_n(x)$ is the complete U-statistic with kernel $h_{n,x}$,
then its first Hoeffding projection dominates:
\[
\frac{U_n(x)-\theta_n(x)-
(s_n/n)\sum_{i=1}^n g_{n,x}(O_i)}{\sigma_n(x)}
\xrightarrow{p}0,
\]
and the triangular-array Lindeberg condition holds for the projected summands.
This is the explicit high-level incrementality condition; it is not asserted for
every adaptive recursive tree \citep{MentchHooker2016,WagerAthey2018}.

\textnormal{(I3)} The number of trees satisfies
$B_n^{-1/2}=o\{\sigma_n(x)\}$.

\textnormal{(I4)} Let
$C_{i,n}(x)=\operatorname{Cov}_\Theta
\{N_i,T_{n,x}\mid\mathcal D_n\}$ be the infinite-tree inclusion covariance and
let $\widehat C_i(x)$ be its empirical version from $B_n$ trees. With
\[
a_{n,s}=\frac{n-1}{n}\left(\frac{n}{n-s_n}\right)^2,
\qquad
q_{i,n}(x)=\frac{s_n}{\sqrt{n(n-1)}}
\{g_{n,x}(O_i)-\overline g_n(x)\},
\]
assume
\[
\sum_{i=1}^n
\{a_{n,s}^{1/2}C_{i,n}(x)-q_{i,n}(x)\}^2
=o_p\{\sigma_n^2(x)\},
\]
\[
a_{n,s}\sum_{i=1}^n
\{\widehat C_i(x)-C_{i,n}(x)\}^2
=o_p\{\sigma_n^2(x)\},
\]
and the sample second moment of $g_{n,x}$ is ratio-consistent for
$\zeta_{1,n}(x)$. The second display is the finite-tree covariance condition;
a crude sufficient condition for bounded trees is
$n/B_n=o\{\sigma_n^2(x)\}$, although less conservative analyses may be
possible.
\end{assumption}

\subsection{Sampling limits and variance estimation}
\label{app:inference_results}

The sampling limit is centered at the repeated-sample forest expectation.
The conditions for changing this centering to tail-survival or cure probability
are stated separately below.

\begin{theorem}[Pointwise sampling limit]
\label{thm:forest_clt}
Under (I1)--(I3),
\[
\{\widehat\psi_{n,B_n}(x)-\theta_n(x)\}/\sigma_n(x)
\xrightarrow{d}N(0,1).
\]
The centering is the repeated-sample expectation of the eligible subsampled
forest. It is not automatically the tail-survival target and is not automatically
the cure probability.
\end{theorem}

\begin{proof}
Let $U_n(x)$ be the complete U-statistic that averages $h_{n,x}$ over all
size-$s_n$ subsamples. Assumption (I2) gives
\[
\frac{U_n(x)-\theta_n(x)}{\sigma_n(x)}
=
\frac{1}{\sqrt{n\zeta_{1,n}(x)}}
\sum_{i=1}^n g_{n,x}(O_i)+o_p(1).
\]
The triangular-array Lindeberg condition yields convergence of the leading sum
to $N(0,1)$. Under one-shot uniform subsampling, the conditional mean of one
randomized tree is exactly $U_n(x)$. Boundedness and conditional independence
give
$\widehat\psi_{n,B_n}(x)-U_n(x)=O_p(B_n^{-1/2})$, which is
$o_p\{\sigma_n(x)\}$ by (I3). Slutsky's theorem proves the result.
\end{proof}

\begin{corollary}[Inference for the tail-survival and cure-probability targets]
\label{cor:forest_clt_cure}
If
\[
\theta_n(x)-\psi_n(x)=o\{\sigma_n(x)\},
\]
then the Gaussian limit in Theorem~\ref{thm:forest_clt} is valid when centered
at the tail-survival target $\psi_n(x)$. It is valid for exact cure only if
additionally
\[
\beta_n(x)=\psi_n(x)-\pi(x)=o\{\sigma_n(x)\}.
\]
Logit-scale inference further requires an interior target and asymptotically
inactive boundary projection.
\end{corollary}

\begin{proof}
Write
\[
\widehat\psi_{n,B_n}(x)-\psi_n(x)
=
\{\widehat\psi_{n,B_n}(x)-\theta_n(x)\}
+
\{\theta_n(x)-\psi_n(x)\}.
\]
The first term divided by $\sigma_n(x)$ converges to $N(0,1)$ and the second is
negligible by assumption. Replacing $\psi_n(x)$ by $\pi(x)$ adds
$\beta_n(x)$, which is negligible on the same scale under the exact-cure bias
condition. The logit statement follows from the delta method only when the
target stays in a compact subset of $(0,1)$ and projection is asymptotically
inactive.
\end{proof}

\begin{theorem}[IJ consistency]
\label{thm:ij_consistency}
Under Theorem~\ref{thm:forest_clt} and (I4),
\[
\widehat V_{\mathrm{IJ}}(x)/\sigma_n^2(x)\xrightarrow{p}1.
\]
Target-centered Wald intervals additionally require the bias conditions in
Corollary~\ref{cor:forest_clt_cure}. Paired contrasts are covered by applying
the same result to the within-tree prediction difference and its corresponding
inclusion covariance.
\end{theorem}

\begin{proof}
For the first-order approximation
$H_{1,b}=\sum_j N_{bj}g_{n,x}(O_j)$, fixed-size sampling gives
\[
\operatorname{Cov}_\Theta(N_{bi},H_{1,b}\mid\mathcal D_n)
=
\frac{s_n(n-s_n)}{n(n-1)}
\{g_{n,x}(O_i)-\overline g_n(x)\}.
\]
Multiplication by
$a_{n,s}^{1/2}=\sqrt{(n-1)/n}\,n/(n-s_n)$ yields exactly $q_{i,n}(x)$.
Assumption (I4) first replaces the infinite-tree inclusion covariance by this
first-order target in aggregate squared error and then replaces the infinite-
tree covariance by its empirical finite-$B_n$ version. Cauchy--Schwarz shows
that both replacements change the sum of squares by
$o_p\{\sigma_n^2(x)\}$. Finally,
\begin{align*}
\frac{1}{\sigma_n^2(x)}\sum_{i=1}^n q_{i,n}^2(x)
&=
\frac{1}{(n-1)\zeta_{1,n}(x)}
\sum_{i=1}^n
\{g_{n,x}(O_i)-\overline g_n(x)\}^2\\
&\xrightarrow{p}1
\end{align*}
by the ratio-consistent second-moment condition. Therefore
$\widehat V_{\mathrm{IJ}}(x)/\sigma_n^2(x)\to_p1$. Applying the same argument
to a within-tree prediction difference proves the paired-contrast statement.
\end{proof}

\noindent\textbf{Proof of Theorem~\ref{thm:inference_summary}}:
\begin{proof}
The Gaussian limit follows from Theorem~\ref{thm:forest_clt}, and the
variance-ratio conclusion follows from Theorem~\ref{thm:ij_consistency}.
Corollary~\ref{cor:forest_clt_cure} gives the two target-centered conclusions
under their respective bias conditions.
\end{proof}

For finite $B$, the software reports the diagnostic correction
\begin{equation}
\widehat V_{\mathrm{IJ},B}^{\mathrm{corr}}(x)
=\left[
\widehat V_{\mathrm{IJ}}(x)
-a_{n,s}\frac{\widehat v_T^{(B)}(x)}{B}
\sum_i\widehat v_{N_i}^{(B)}
\right]_+ .
\label{eq:finite_b_ij_correction}
\end{equation}
This product-of-marginal-variances subtraction is not implied by
Theorem~\ref{thm:ij_consistency} \citep{WagerHastieEfron2014}. It may exhaust
the raw variance when $B$ is too small and cannot remove localization,
regularization, nuisance-estimation, or residual-tail bias. The finite-$B$
correction is therefore treated as a diagnostic, and the simulations evaluate
variance calibration around the forest expectation separately from coverage of
the scientific target.

\clearpage
\setcounter{section}{0}
\setcounter{table}{0}
\setcounter{figure}{0}
\setcounter{equation}{0}
\renewcommand{\thesection}{Supplementary Appendix \Alph{section}}
\renewcommand{\thesubsection}{\Alph{section}.\arabic{subsection}}
\renewcommand{\tablename}{Supplementary Table}
\renewcommand{\thetable}{\arabic{table}}
\renewcommand{\figurename}{Supplementary Figure}
\renewcommand{\thefigure}{\arabic{figure}}
\renewcommand{\theequation}{\Alph{section}.\arabic{equation}}

\section*{Supplementary Material}

This supplementary material contains simulation details, registry specifications,
additional diagnostics, and an exploratory application.

\section{Simulation designs and results}
\label{supp:simulation}

\subsection{Objectives and reporting strategy}
\label{supp:simulation_objectives}

The Supplementary Material gives the complete simulation designs,
sensitivity analyses, FOSP implementation details, and additional estimator
formulas. Technical proofs are in Appendix~A of the article.

The primary questions concern recovery of the cure variable, estimation of
\(\pi(x)\), stopping under cure homogeneity, and the effects of tree depth,
follow-up maturity, and censoring. We pair cure-probability MSE, latent-cure
Brier score, and AUC with the proportions of root splits on cure, latency, and
noise variables. Root stopping is evaluated separately. These measures
distinguish prediction from recovery of the incidence structure. For
inference, calibration around the repeated-sample forest expectation is
reported separately from coverage of the tail-survival and exact cure
contrasts.

The main and depth designs comprised 2,400 and 1,200 generated data sets,
respectively, with three methods fitted to each. The null design comprised
2,400 threshold-specific CBST-forest fits and a threshold-invariant log-rank
reference. The four inference profiles comprised 5,600 forests, and the
calibration pilot added 720. These counts distinguish generated data sets from
fitted forests; all reported fits completed without computational failures.

\subsection{Cure--latency separation design}
\label{supp:simulation_design}
\label{supp:simulation_main_results}

For each subject, we generated
\[
X_1\sim \operatorname{Uniform}(0,1),\qquad
X_2\sim \operatorname{Bernoulli}(0.5),\qquad
X_3\sim N(0,1),
\]
independently. The covariate \(X_1\) determines the cure probability,
\[
\pi_0(X_1)=\operatorname{logit}^{-1}\{-0.55+a(X_1-0.5)\},
\qquad a\in\{1.5,3.0\},
\]
and \(Z\sim\operatorname{Bernoulli}\{\pi_0(X_1)\}\), where \(Z=1\)
denotes a cured subject. The covariate \(X_2\) has no effect on
\(\pi_0\), but determines the susceptible event-time distribution. For
\(Z=0\), we generated
\[
T=8U^{1/q(X_2)},\qquad U\sim\operatorname{Uniform}(0,1),
\]
where \(q(1)=3^b\), \(q(0)=0.75^b\), and
\(b\in\{0.8,1.2\}\). Thus, \(X_2\) can produce substantial early
survival separation without changing the long-term cure fraction. We set
\(T=\infty\) when \(Z=1\), and \(X_3\) was a noise covariate.

Two censoring mechanisms were considered. Under marginally independent
censoring, \(C=12V^{1/1.5}\), where
\(V\sim\operatorname{Uniform}(0,1)\). Under covariate-dependent censoring,
\[
C=12V^{1/k(X_1)},\qquad k(X_1)=3.0-2.3X_1.
\]
The latter mechanism is independent of the event time conditional on the
covariates, but it is not independent within an unadjusted daughter node. The
observed outcome is \(\widetilde T=\min(T,C)\) with event indicator
\(\Delta=I(T\leq C)\).

We compared three forests under shared global controls: the proposed KM
CBST forest, an oracle-IPCW CBST forest using the known
censoring survival function, and a log-rank forest with KM terminal-node
estimation. All forests used 300 trees, 80\% subsampling, honest 50--50
structure and estimation splits, three candidate variables at each node, nine
candidate cutpoints, minimum daughter size, maximum depth two, and the common
tail interval \([8.2,10]\). Candidates evaluated with the cure-fraction based
splitting criterion additionally had to satisfy the event and late-risk-set
maturity filters; log-rank candidates did not. The
training sizes were \(n=500,1000,2000\), and each fitted
forest was evaluated in an independently generated test sample of size 5000.
For every combination of \(n\), cure strength, latency strength, and censoring
mechanism, 100 Monte Carlo repetitions were performed. Methods were fitted to
the same simulated data in each repetition. The oracle-IPCW procedure uses the
known censoring survival and therefore does not introduce a fitted nuisance
outside the tree kernel.

The main results and their interpretation are reported in
Section~\ref{sec:simulation}, Table~\ref{tab:main_confirmatory_results}.
The following analyses examine sensitivity and inference.

\subsection{Sensitivity to tree depth}
\label{sec:simulation_depth}

The depth experiment used \(n=1000\), 200 trees, depths one to three, an
independent test sample of size 5000, and 50 repetitions for each combination
of cure strength, latency strength, and censoring mechanism. Thus, each table
entry averages 400 data sets. Increasing depth from one to two reduced the KM
CBST-forest MSE by 24.4\%, from 0.00458 to 0.00346. Depth three produced no
material further improvement and increased the mean computation time. We
therefore use depth two as the primary shallow-forest specification.

\begin{table}[!htbp]
\centering
\scriptsize
\caption{Sensitivity to maximum tree depth. Results average over two cure
strengths, two latency strengths, two censoring mechanisms, and 50 repetitions
per scenario at \(n=1000\). Time is the mean number of seconds per fitted
forest on the simulation computer.}
\label{tab:depth_sensitivity_results}
\begin{tabular}{@{}rlrrr@{}}
\toprule
Depth & Method & MSE\(_\pi\) & AUC & Time \\
\midrule
1 & CBST KM     & 0.00458 & 0.6641 & 5.02 \\
  & CBST IPCW   & 0.00687 & 0.6592 & 2.82 \\
  & Log-rank KM & 0.00866 & 0.6551 & 3.64 \\
\addlinespace
2 & CBST KM     & 0.00346 & 0.6660 & 9.13 \\
  & CBST IPCW   & 0.00588 & 0.6602 & 5.29 \\
  & Log-rank KM & 0.00646 & 0.6604 & 7.64 \\
\addlinespace
3 & CBST KM     & 0.00346 & 0.6659 & 10.30 \\
  & CBST IPCW   & 0.00596 & 0.6600 & 6.15 \\
  & Log-rank KM & 0.00624 & 0.6599 & 12.15 \\
\bottomrule
\end{tabular}
\end{table}

\subsection{Forest inference}
\label{supp:simulation_inference}

The inference experiment was designed separately because asymptotic
normality concerns an honest subsampled forest at fixed covariate profiles,
rather than test-set prediction averaged over a population. We generated
\(X_1\sim\operatorname{Uniform}(-1,1)\),
\(X_2\sim\operatorname{Bernoulli}(0.5)\), and \(X_3\sim N(0,1)\). In the
completed core profile, the cure probability was
\[
\pi_0(x_1)=
\begin{cases}
0.18, & x_1\leq -0.35,\\
0.40, & -0.35<x_1\leq 0.35,\\
0.68, & x_1>0.35.
\end{cases}
\]
Susceptible event times followed a Weibull distribution with shape 1.7 and
scale
\[
\ell(x)=\ell_0\exp(0.25x_2-0.08x_1),
\]
where \(\ell_0=2.0,3.5,5.0\) defined mature, limited, and immature follow-up,
respectively. Independent censoring was generated on \([0,13]\), and the
forest target used a common five-point tail grid on \([8,10]\).

We considered the low, middle, and high profiles
\[
(x_1,x_2,x_3)\in\{(-0.7,0,0),(0,0,0),(0.7,0,0)\},
\]
with particular interest in the high-minus-low contrast. Each forest used
2000 trees, subsampling without
replacement of size
\(s_n=\min\{n-1,\max(100,\lfloor n^{0.9}\rfloor)\}\), honest structure and
estimation samples, and maximum depth two. For each combination of
\(n\in\{500,1000,2000\}\) and follow-up maturity, 300 repetitions were
performed. The analysis distinguished the exact cure contrast
\(\Delta_\pi\) from the observable tail-survival contrast
\(\Delta_\psi\) through the reported tail-bias column. Interval coverage in
this profile is reported for \(\Delta_\psi\) and for the repeated-sample
forest expectation. Exact-cure coverage is evaluated in the independent
maturity stress test below.

Table~\ref{tab:inference_core_results} separates tail-survival estimation error
from residual-tail bias. Immature follow-up produced the largest difference
between \(\Delta_\pi\) and \(\Delta_\psi\), as predicted by the theoretical
decomposition.

\begin{table}[!htbp]
\centering
\scriptsize
\caption{Inference for the high-minus-low CBST-forest contrast in the
completed step-function core profile (300 repetitions per row and 2000 trees
per forest). Tail bias is \(\Delta_\psi-\Delta_\pi\); Bias\(_\psi\) is the
mean estimate minus \(\Delta_\psi\); SD is the empirical Monte Carlo standard
deviation. IJ/SD compares the finite-\(B\)-corrected IJ standard error with SD.
The reported IJ uses the without-replacement factor
\(a_{n,s}\). Target and centered coverage refer, respectively, to
\(\Delta_\psi\) and to the repeated-sample mean forest prediction.}
\label{tab:inference_core_results}
\begin{tabular}{@{}rllrrrrr@{}}
\toprule
\(n\) & Follow-up & Tail bias & Bias\(_\psi\) & SD & IJ/SD &
Target cov. & Centered cov. \\
\midrule
500  & Immature & -0.0553 & -0.1290 & 0.0535 & 1.018 & 0.317 & 0.937 \\
     & Limited  & -0.0087 & -0.1106 & 0.0521 & 0.971 & 0.447 & 0.920 \\
     & Mature   &  0.0000 & -0.0991 & 0.0493 & 0.998 & 0.510 & 0.930 \\
\addlinespace
1000 & Immature & -0.0553 & -0.0911 & 0.0424 & 1.041 & 0.463 & 0.957 \\
     & Limited  & -0.0087 & -0.0633 & 0.0412 & 1.028 & 0.680 & 0.953 \\
     & Mature   &  0.0000 & -0.0431 & 0.0396 & 1.012 & 0.840 & 0.943 \\
\addlinespace
2000 & Immature & -0.0553 & -0.0645 & 0.0302 & 1.059 & 0.463 & 0.960 \\
     & Limited  & -0.0087 & -0.0382 & 0.0281 & 1.043 & 0.753 & 0.953 \\
     & Mature   &  0.0000 & -0.0178 & 0.0259 & 1.041 & 0.913 & 0.957 \\
\bottomrule
\end{tabular}
\end{table}

The variance and coverage columns answer different questions. After correcting
for fixed-size subsampling, the estimated standard errors closely reproduced
the empirical sampling variation: IJ/SD ranged from 0.971 to 1.059, and
centered coverage ranged from 0.920 to 0.960. Coverage of \(\Delta_\psi\) was
lower because the forest bias was large relative to its standard deviation.
Under mature follow-up, target coverage increased from 0.510 at \(n=500\) to
0.913 at \(n=2000\), as the absolute bias decreased from 0.0991 to 0.0178.
Thus, the variance estimator supports the Gaussian approximation around the
forest expectation, while inference for the tail-survival or cure-probability target also
requires the bias condition in Corollary~\ref{cor:forest_clt_cure}.

\subsection{Additional inference diagnostics}
\label{sec:simulation_additional_inference}

The smooth profile replaces the step cure probability by
\[
\pi_0(x_1)=\operatorname{logit}^{-1}\{-0.45+1.65x_1
+0.35\sin(\pi x_1)\}
\]
and uses 1500 trees and 200 repetitions for each combination of
\(n\in\{1000,2000\}\) and limited or mature follow-up. All 800 prespecified
forests completed without failed fits, nonfinite estimates, or zero corrected
IJ standard errors.

\begin{table}[!htbp]
\centering
\scriptsize
\caption{Inference for the high-minus-low contrast under a smooth cure
probability (200 repetitions per row and 1500 trees per forest).
Bias\(_\psi\) is the mean estimate minus the tail-survival contrast; SD is the
empirical Monte Carlo standard deviation. IJ/SD uses the finite-\(B\) and
without-replacement corrections. Target and centered coverage refer to nominal
95\% intervals for the tail-survival contrast and the repeated-sample mean,
respectively.}
\label{tab:inference_smooth_results}
\begin{tabular}{@{}rlrrrrr@{}}
\toprule
\(n\) & Follow-up & Bias\(_\psi\) & SD & IJ/SD & Target cov. & Centered cov. \\
\midrule
1000 & Limited & -0.0637 & 0.0387 & 0.964 & 0.595 & 0.930 \\
     & Mature  & -0.0501 & 0.0345 & 1.026 & 0.755 & 0.940 \\
2000 & Limited & -0.0405 & 0.0293 & 1.009 & 0.760 & 0.935 \\
     & Mature  & -0.0238 & 0.0269 & 1.059 & 0.870 & 0.970 \\
\bottomrule
\end{tabular}
\end{table}

The smooth-incidence results reproduce the main diagnosis of the step-function
experiment. The absolute tail-survival bias decreased from 0.0637 to 0.0405
under limited follow-up and from 0.0501 to 0.0238 under mature follow-up when
the sample size doubled. In the mature design, the exact cure and tail-survival
contrasts differed by less than 0.00002, so the remaining bias cannot be
attributed to an immature plateau. It is instead consistent with finite-sample
forest localization and regularization bias.

The corrected IJ standard error was also well calibrated under smooth
incidence: IJ/SD ranged from 0.964 to 1.059, and centered coverage ranged from
0.930 to 0.970. Target coverage increased from 0.595--0.755 at \(n=1000\) to
0.760--0.870 at \(n=2000\), following the reduction in point-estimation bias.
Because the variance calibration is similar for step and smooth incidence
functions, non-smoothness is not the principal explanation for the remaining
target undercoverage.

The censoring profile used the step cure model and covariate-dependent
censoring. It compared unadjusted KM estimation with oracle IPCW, keeping the
censoring weights fixed within each simulated data set so that the forest IJ
continued to refer to the one-shot tree kernel. Each combination of sample
size, follow-up maturity, and method was evaluated in 200 repetitions using
1500 trees. The observed censoring fraction was approximately 49\% under
limited follow-up and 45\% under mature follow-up. All 1,600 prespecified
forests completed without failed fits, missing tasks, or nonfinite estimates.

\begin{table}[!htbp]
\centering
\scriptsize
\caption{High-minus-low contrast under covariate-dependent censoring (200
repetitions per row and 1500 trees per forest). IJ/SD uses the finite-\(B\) and
without-replacement corrections. Target and centered coverage distinguish the
tail-survival contrast from the mean forest prediction. FB is the mean terminal-
node fallback fraction.}
\label{tab:inference_censoring_results}
\resizebox{\textwidth}{!}{%
\begin{tabular}{@{}rllrrrrrr@{}}
\toprule
\(n\) & Follow-up & Method & Bias\(_\psi\) & RMSE\(_\psi\) & IJ/SD &
Target cov. & Centered cov. & FB \\
\midrule
1000 & Limited & KM          & -0.0700 & 0.0812 & 1.026 & 0.635 & 0.965 & 0.037 \\
     &         & Oracle IPCW & -0.1092 & 0.1230 & 1.024 & 0.540 & 0.950 & 0.045 \\
     & Mature  & KM          & -0.0516 & 0.0637 & 1.056 & 0.775 & 0.950 & 0.038 \\
     &         & Oracle IPCW & -0.0912 & 0.1100 & 0.952 & 0.645 & 0.945 & 0.047 \\
\addlinespace
2000 & Limited & KM          & -0.0351 & 0.0461 & 1.008 & 0.820 & 0.935 & 0.008 \\
     &         & Oracle IPCW & -0.0623 & 0.0782 & 0.969 & 0.715 & 0.940 & 0.011 \\
     & Mature  & KM          & -0.0249 & 0.0368 & 1.020 & 0.880 & 0.940 & 0.011 \\
     &         & Oracle IPCW & -0.0539 & 0.0693 & 1.033 & 0.775 & 0.935 & 0.013 \\
\bottomrule
\end{tabular}}
\end{table}

The KM forest had the smallest absolute bias and RMSE in all four
sample-size--maturity combinations. Under limited follow-up and \(n=1000\),
for example, its RMSE was 0.0812, compared with 0.1230 for oracle IPCW. Under
mature follow-up and \(n=2000\), the corresponding values were 0.0368 and
0.0693. Thus, weighting did not improve
finite-sample estimation in this design, even when the censoring distribution
was known. The IPCW forests changed both split selection and terminal-node
estimation and incurred additional variability, whereas the shallow local
partition made the unadjusted KM estimator comparatively stable. This finding
is specific to the censoring mechanism and forest architecture considered and
does not justify omitting censoring adjustment when nodewise independent
censoring is implausible. For every method, bias and fallback frequency
decreased as the sample size increased. IJ/SD remained between 0.952 and
1.056, and centered coverage remained between 0.935 and 0.965. These results
support the variance scale around the method-specific forest expectation;
target coverage also reflects the corresponding forest bias.

Finally, the tree-number profile fixed \(n=1000\), mature follow-up, and the KM
forest, and varied \(B\in\{250,500,1000,2000,4000\}\). One hundred repetitions
were performed for each value of \(B\). This profile directly evaluates the
finite-\(B\) adjustment and the conditional Monte Carlo component.

\begin{table}[!htbp]
\centering
\scriptsize
\caption{Sensitivity of high-minus-low contrast inference to the number of
trees at \(n=1000\) under mature follow-up (100 repetitions per row). MC SE is
the conditional finite-forest Monte Carlo standard error; correction is the
median fraction of raw IJ variance removed by the finite-\(B\) adjustment. IJ
SE and IJ/SD also include the without-replacement factor.}
\label{tab:inference_b_results}
\resizebox{\textwidth}{!}{%
\begin{tabular}{@{}rrrrrrr@{}}
\toprule
\(B\) & Correction & IJ SE & MC SE & IJ/SD & Target cov. & Centered cov. \\
\midrule
250  & 0.947 & 0.0367 & 0.0052 & 0.955 & 0.750 & 0.840 \\
500  & 0.893 & 0.0392 & 0.0037 & 0.987 & 0.850 & 0.930 \\
1000 & 0.808 & 0.0398 & 0.0026 & 1.009 & 0.900 & 0.920 \\
2000 & 0.681 & 0.0396 & 0.0018 & 0.997 & 0.910 & 0.920 \\
4000 & 0.513 & 0.0396 & 0.0013 & 1.008 & 0.920 & 0.940 \\
\bottomrule
\end{tabular}}
\end{table}

Across \(B\), tail-survival bias remained between -0.0362 and -0.0370 and the
empirical standard deviation between 0.0384 and 0.0397. The conditional Monte
Carlo standard error decreased from 0.0052 to 0.0013. After both corrections,
IJ/SD remained near one, but centered coverage was only 0.840 with 250 trees,
increasing to 0.920--0.940 with at least 500 trees. Thus, a reasonable average
standard error did not ensure stable interval coverage at small \(B\);
target coverage also remained affected by bias.

The without-replacement factor addresses a separate issue: at \(n=1000\),
\(s_n=\lfloor n^{0.9}\rfloor\) samples approximately half the observations,
attenuating the unadjusted inclusion covariance. The results support the
corrected variance scale as an empirical diagnostic, while the finite-\(B\)
subtraction remains an approximation. They do not establish the CLT or
remove the bias conditions required for target inference.

\subsection{Without-replacement calibration pilot}
\label{supp:ij_fpc_pilot}

We verified the implementation in an additional calibration pilot at
\(n=1000\), under mature follow-up and independent censoring, for step and
smooth incidence functions. For each structure, nine prespecified
configurations were drawn from sampling fractions 0.25, 0.35, and 0.50,
maximum depths two to four, and 12, 40, or 100 candidate cutpoints; the wording
does not imply a full 27-cell factorial. Forty paired repetitions were reused
across configurations, with 1000 trees per forest, giving 720 fits without
failures.

Table~\ref{tab:ij_fpc_pilot} reports the high-minus-low contrast for the
lowest-score configuration within each sampling fraction and structure. The
prespecified score combined standardized absolute bias, IJ calibration,
centered coverage, fallback frequency, and a four-group cross-product IJ
stability
measure. Because the displayed configurations were selected within this pilot,
they are calibration diagnostics rather than an independent confirmatory
comparison.

The cross-product calculation partitions the \(B=1000\) already fitted trees
cyclically into four groups; it does not split observations or refit forests.
Within group \(g\), both inclusion indicators and tree predictions are centered
by their group means and multiplied by the same without-replacement scale as
the analytic IJ, producing group-specific influence estimates
\(\widehat C_{ig}(x)\). For each of the six unordered pairs \((g,h)\), the
profile variance contribution is
\(\sum_i\widehat C_{ig}(x)\widehat C_{ih}(x)\); the reported estimate averages
these six cross-products. The contrast version replaces each influence by its
high-minus-low difference before multiplication. A negative cross-product
variance produces a missing standard error rather than truncation at zero, and
the fraction of negative estimates is retained as a stability diagnostic.
CF/SD is the mean valid cross-product standard error divided by the empirical
Monte Carlo standard deviation.

For each structure and tuning configuration, the ranking averages over four
estimands: the cure prediction at \(x_1=-0.70,0,0.70\), with
\(x_2=x_3=0\), and the high-minus-low contrast. Its exact score is
\[
\begin{aligned}
\mathrm{score}={}&\overline{|\mathrm{bias}|/\mathrm{SD}}
+1.25\,\overline{|\log(\mathrm{IJ/SD})|}
+0.75\,\overline{|\log(\mathrm{CF/SD})|}\\
&+2\,\overline{|\mathrm{coverage}_{c}-0.95|}
+0.5\,\overline{\mathrm{fallback}}
+0.25\,\overline{\mathrm{negative\ CF}}.
\end{aligned}
\]
where bars denote finite averages across the four estimands and
\(\mathrm{coverage}_{c}\) is coverage around the repeated-sample forest mean.
For the leading configuration within each structure, the prespecified checks
were mean analytic calibration error at most \(|\log(1.15)|\), mean centered-
coverage error at most 0.04, mean negative cross-product fraction at most 0.10,
and mean absolute bias no greater than 0.50 empirical standard deviations.

\begin{table}[!htbp]
\centering
\scriptsize
\caption{Without-replacement IJ calibration pilot for the high-minus-low
contrast (40 paired repetitions and 1000 trees per configuration). Fraction is
the tree subsampling fraction. Bias is relative to the tail-survival target.
IJ/SD and CF/SD compare the analytic and four-group cross-product IJ standard
errors with the
empirical standard deviation. Target and centered coverage refer to nominal
95\% intervals.}
\label{tab:ij_fpc_pilot}
\begin{tabular}{@{}lrrrrrrr@{}}
\toprule
Structure & Fraction & Depth & Bias\(_\psi\) & IJ/SD & CF/SD &
Target cov. & Centered cov. \\
\midrule
Smooth & 0.25 & 4 & -0.1199 & 1.034 & 1.022 & 0.000 & 0.925 \\
       & 0.35 & 3 & -0.0817 & 1.047 & 1.039 & 0.275 & 0.975 \\
       & 0.50 & 4 & -0.0541 & 1.074 & 1.028 & 0.625 & 0.950 \\
\addlinespace
Step   & 0.25 & 4 & -0.1030 & 1.039 & 1.022 & 0.075 & 0.975 \\
       & 0.35 & 4 & -0.0657 & 1.108 & 1.089 & 0.475 & 0.975 \\
       & 0.50 & 3 & -0.0394 & 1.141 & 1.184 & 0.800 & 0.925 \\
\bottomrule
\end{tabular}
\end{table}

Across the six displayed rows, IJ/SD ranged from 1.034 to 1.141 and
centered coverage from 0.925 to 0.975. The reported cross-product ratios were close
to the analytic values except in the final displayed row. Negative cross-product
variance occurred in 2.5\% of the step-function contrasts for fraction 0.50
and depth three and was absent in the other displayed rows. A larger sampling
fraction reduced, but did not eliminate, target bias. The aggregate
summary reports mean absolute biases of 1.102 and 0.695 empirical standard
deviations for the leading smooth and step configurations, both exceeding the
prespecified threshold of 0.50 under the four-estimand aggregation defined
above. The pilot supports the variance correction only
around the forest expectation and retains the need for explicit bias control
for target inference.

\subsection{Independent bias, maturity, and censoring stress tests}
\label{supp:bias_strengthening}

The configurations selected in the calibration pilot were next frozen and
evaluated with independent random-number streams. For each structure, 300
replications were performed at \(n=1000\) and \(n=2000\), with 1000 trees,
65\% subsampling, mature follow-up, and independent censoring. This experiment
was designed to verify the direction of point-estimation bias rather than to
select another configuration.

\begin{table}[!htbp]
\centering
\scriptsize
\caption[Independent confirmation of the locked smooth-incidence contrast.]
{Independent confirmation of the high-minus-low contrast for the
locked smooth and step configurations (300 replications per row). Bias is
relative to the exact cure contrast. Zero IJ is the fraction for which the
finite-\(B\) subtraction exhausted the raw IJ variance.}
\label{tab:bias_independent_confirmation}
\begin{tabular}{@{}lrrrrrrr@{}}
\toprule
Structure & \(n\) & Bias & SD & IJ/SD & Centered cov. & Cure cov. & Zero IJ \\
\midrule
Smooth & 1000 & -0.0378 & 0.0396 & 0.982 & 0.923 & 0.843 & 0.007 \\
       & 2000 & -0.0202 & 0.0299 & 0.979 & 0.863 & 0.817 & 0.053 \\
Step   & 1000 & -0.0212 & 0.0392 & 1.077 & 0.943 & 0.923 & 0.003 \\
       & 2000 & -0.0101 & 0.0282 & 1.054 & 0.863 & 0.873 & 0.050 \\
\bottomrule
\end{tabular}
\end{table}

The absolute bias decreased when \(n\) doubled for both structures. The
decomposition attributed most of the remaining bias to localization: its mean
contribution changed from -0.0285 to -0.0162 for smooth incidence and from
-0.0184 to -0.0076 for step incidence, while the residual-tail component was
smaller than \(2\times10^{-5}\). Thus, the independent experiment confirms
that point-estimation bias decreases, but the smooth configuration still
requires additional undersmoothing before unconditional cure-target coverage
can be claimed.

The \(n=2000\) rows also identify a finite-forest limitation. The median
finite-\(B\) subtraction removed approximately 95\% of the raw IJ variance,
compared with 89\% at \(n=1000\), and approximately 5\% of corrected variances
were truncated at zero. Consequently, a near-unit ratio of mean IJ standard
error to SD did not produce nominal centered coverage. The corresponding raw
IJ intervals covered essentially every target because their standard errors
were much too large. This experiment therefore does not overturn the
theory-aligned and tree-number results above; it shows that \(B=1000\) is
insufficient for stable covariance recovery in the more aggressive locked
configuration at \(n=2000\). No target-level interval from this configuration
is recommended.

We next held \(n=1000\) fixed and varied the susceptible-time scale to create
mature, limited, and immature tails. One hundred independent replications were
performed for each structure and maturity level.
Table~\ref{tab:tail_maturity_strengthening} separates the design-level difference
\(\Delta_\psi-\Delta_\pi\) from the forest's bias for the observable
tail-survival target.

\begin{table}[!htbp]
\centering
\scriptsize
\caption{Tail-maturity stress test (100 replications per row). Tail bias is
\(\Delta_\psi-\Delta_\pi\), and Bias\(_\psi\) is the forest bias relative to
the tail-survival contrast. The final columns report nominal 95\% coverage for
the exact cure and tail-survival contrasts.}
\label{tab:tail_maturity_strengthening}
\begin{tabular}{@{}llrrrr@{}}
\toprule
Structure & Follow-up & Tail bias & Bias\(_\psi\) & Cure cov. & Tail cov. \\
\midrule
Smooth & Immature & -0.0622 & -0.0665 & 0.18 & 0.71 \\
       & Limited  & -0.0096 & -0.0456 & 0.69 & 0.76 \\
       & Mature   &  0.0000 & -0.0396 & 0.84 & 0.84 \\
Step   & Immature & -0.0553 & -0.0664 & 0.23 & 0.70 \\
       & Limited  & -0.0087 & -0.0482 & 0.77 & 0.80 \\
       & Mature   &  0.0000 & -0.0219 & 0.93 & 0.93 \\
\bottomrule
\end{tabular}
\end{table}

Under immature follow-up, coverage of the exact cure contrast fell to
0.18--0.23, whereas coverage of the observable tail contrast was 0.70--0.71.
Maturing the tail removed the design-level discrepancy but not the forest
localization bias, especially for smooth incidence. In every maturity cell,
however, more than 99.9\% of tree roots selected \(X_1\). The structural
signal remained identifiable even when exact cure was not.

Finally, a censoring stress test compared marginal KM, oracle IPCW, and
fivefold Cox-IPCW under covariate-dependent censoring, using 100 independent
replications for each structure. The independent-censoring KM row is included
as a reference. All methods selected \(X_1\) at more than 99.7\% of roots.

\begin{table}[!htbp]
\centering
\scriptsize
\caption{Censoring stress test for the high-minus-low contrast at \(n=1000\)
(100 replications per row). Terminal and localization are the corresponding
mean components of the bias decomposition.}
\label{tab:censoring_strengthening}
\begin{tabular}{@{}lllrrrr@{}}
\toprule
Structure & Censoring & Estimator & Bias & RMSE & Terminal & Localization \\
\midrule
Smooth & Independent & KM          & -0.0332 & 0.0485 & -0.0058 & -0.0274 \\
       & Covariate   & KM          & -0.0499 & 0.0621 & -0.0171 & -0.0328 \\
       & Covariate   & Oracle IPCW & -0.0599 & 0.0812 & -0.0099 & -0.0500 \\
       & Covariate   & Cox IPCW    & -0.0425 & 0.0673 &  0.0066 & -0.0491 \\
\addlinespace
Step   & Independent & KM          & -0.0238 & 0.0431 & -0.0053 & -0.0184 \\
       & Covariate   & KM          & -0.0312 & 0.0503 & -0.0085 & -0.0226 \\
       & Covariate   & Oracle IPCW & -0.0427 & 0.0720 &  0.0014 & -0.0442 \\
       & Covariate   & Cox IPCW    & -0.0234 & 0.0570 &  0.0198 & -0.0432 \\
\bottomrule
\end{tabular}
\end{table}

Dependent censoring increased the KM error relative to its
independent-censoring reference. IPCW reduced or reversed the terminal
component, but it also changed the fitted partitions, increased variability,
and produced a more negative localization component. Cox-IPCW had the smallest
total bias in the step design partly because its positive terminal and negative
localization components canceled. Hence, this experiment supports KM when its
nodewise censoring assumption is plausible and IPCW as a sensitivity analysis;
it does not establish uniform superiority of weighting. Because the
Cox nuisance was shared across trees, its ordinary IJ columns remain
descriptive.

\subsection{Repeated nonlinear benchmark}
\label{sec:simulation_complex}

We also considered a multivariable design in which incidence and susceptible
latency were nonlinear. Ten correlated Gaussian covariates were generated with
pairwise correlation 0.30. The cure linear predictor contained a sinusoidal
effect, an interaction, a threshold, and a quadratic term,
\[
\operatorname{logit}\{\pi_0(x)\}
=-0.60+1.10\sin(x_1)+0.75x_2x_3
+0.70I(x_4>0)-0.25x_8^2.
\]
The susceptible event time was
\(10U^{1/q(x)}\), where
\(q(x)=\exp(0.45x_5-0.28x_6^2+0.22x_7)\). We used both independent and
covariate-dependent censoring, training sizes \(n=500,1000,2000\), and an
independent test sample of size 3000. For every sample-size--censoring cell,
200 paired replications were performed with 300 trees.

The primary comparison is with a log-rank forest under shared global controls.
Candidates evaluated with the cure-fraction based splitting criterion
additionally satisfy tail-maturity admissibility, whereas
log-rank candidates do not. We also included two external references: a 300-tree
standard random survival
forest fitted with \texttt{randomForestSRC}, and a proportional-hazards
mixture-cure model fitted with \texttt{smcure}. The latter entered all ten
covariates linearly in both incidence and latency, so it was intentionally
misspecified for the nonlinear truth. Table~\ref{tab:complex_simulation_results}
averages equally over the two censoring mechanisms. Because the susceptible
distribution has support ending at 10 years, survival at that endpoint equals
the exact cure probability. All 4,800 method fits completed successfully.

\begin{table}[!htbp]
\centering
\scriptsize
\caption{Repeated test-set performance under nonlinear incidence and
susceptible latency (400 paired replications per sample size after averaging
over the two censoring mechanisms). Brier and AUC use latent cure status.
MSE\(_\psi\) evaluates survival at 8.5 years using the tree-first
coherent aggregation.}
\label{tab:complex_simulation_results}
\begin{tabular}{@{}rlrrrr@{}}
\toprule
\(n\) & Forest & MSE\(_\pi\) & Brier & AUC & MSE\(_\psi\) \\
\midrule
500  & CBST & 0.03797 & 0.23172 & 0.6898 & 0.02771 \\
     & Log-rank      & 0.03962 & 0.23336 & 0.6900 & 0.02795 \\
     & Standard RSF  & 0.02638 & 0.22003 & 0.6934 & 0.02068 \\
     & Mixture cure  & 0.03388 & 0.22752 & 0.6728 & 0.03422 \\
1000 & CBST & 0.03236 & 0.22611 & 0.6994 & 0.02322 \\
     & Log-rank      & 0.03372 & 0.22747 & 0.6983 & 0.02372 \\
     & Standard RSF  & 0.02214 & 0.21580 & 0.7064 & 0.01714 \\
     & Mixture cure  & 0.03014 & 0.22382 & 0.6827 & 0.03015 \\
2000 & CBST & 0.02837 & 0.22217 & 0.7071 & 0.02020 \\
     & Log-rank      & 0.02966 & 0.22347 & 0.7036 & 0.02094 \\
     & Standard RSF  & 0.01899 & 0.21278 & 0.7160 & 0.01457 \\
     & Mixture cure  & 0.02823 & 0.22205 & 0.6887 & 0.02825 \\
\bottomrule
\end{tabular}
\end{table}

The proposed forest reduced cure-probability MSE by 4.0--4.3\% relative to the
log-rank comparator in this more diffuse nonlinear design. Its AUC was
essentially unchanged at \(n=500\) and modestly higher at \(n=1000\) and
\(n=2000\). With coherent tree-first aggregation, its 8.5-year survival MSE was
also lower than that of the log-rank comparator, by 0.9\%, 2.1\%, and 3.5\%
for \(n=500,1000,2000\), respectively. This finite-time advantage was modest at
the smallest sample size but became more pronounced as the sample size
increased. Cure-probability MSE, latent-cure Brier score, and AUC were unchanged
by the aggregation correction.

The external references make that boundary more explicit. Standard RSF had
the smallest cure-probability MSE, latent-status Brier score, and 8.5-year
survival MSE, and the largest AUC, at every sample size. The misspecified
mixture-cure model had slightly smaller cure MSE and Brier score than the
CBST forest, but appreciably lower AUC and larger 8.5-year survival
MSE. Thus the proposed method is not a universally optimal prediction engine.
Its contribution is a tail-supported CBST partition whose behavior
can be compared with the log-rank partition under shared global controls; the
comparison evaluates the complete partitioning rules rather than isolated
score numerators. When generic
prediction is the only goal, a standard RSF can be preferable.

\subsection{FOSP implementation details}
\label{supp:fosp_controls}

Supplementary Table~\ref{tab:supp_fosp_flow} gives the sequential registry checks applied
before restricting diagnosis years and selecting the two case-study cancer
groups.

\begin{table}[!htbp]
\centering
\small
\caption{FOSP registry flow before the temporal case-study restrictions.
Exclusions are sequential.}
\label{tab:supp_fosp_flow}
\begin{tabular}{lrr}
\toprule
Step & Excluded & Remaining \\
\midrule
Raw registry snapshot & --- & 1,257,217 \\
Nonfinite follow-up time & 1 & 1,257,216 \\
Nonpositive follow-up time & 34,781 & 1,222,435 \\
Missing or invalid status, topography, or age & 0 & 1,222,435 \\
Age greater than 110 years & 1 & 1,222,434 \\
\bottomrule
\end{tabular}
\end{table}

All preprocessing was learned within the 2000--2009 training cohort.
Nonfinite numeric values were replaced by the training median; missing
categorical values received an explicit \texttt{unknown} level; rare categories
were collapsed to \texttt{other}; and unseen confirmation levels were mapped
to the training \texttt{other} level or, if unavailable, the modal training
level. Registry administrative codes such as stage X/Y, grade 8/9, and basis
of diagnosis 9 were retained as factor levels rather than treated as clinical
measurements.

Each CBST and log-rank forest used 1000 trees, 80\% subsampling
without replacement, disjoint 50--50 structure and estimation sets, maximum
depth five, \(m_{\mathrm{try}}=\lfloor\sqrt p\rfloor\), and 20
empirical-quantile cutpoints per selected variable. The tail grid contained 24 equally weighted
points over years 8--10. For training size \(n\), the forest controls were
\[
n_{\min}=\max\{35,\min(120,\lfloor\sqrt n/2\rfloor)\},\quad
r_8=\max\{20,\min(80,\lfloor0.003n\rfloor)\},
\]
\[
r_{10}=\max\{8,\min(35,\lfloor0.0015n\rfloor)\},\quad
d_{<8}=\max\{8,\min(35,\lfloor0.002n\rfloor)\}.
\]
Here \(r_8,r_{10}\) are minimum risk-set sizes and \(d_{<8}\) is the minimum
number of events before the tail. These maturity controls were applied during
candidate evaluation with the cure-fraction based splitting criterion;
log-rank candidate scoring enforced minimum
daughter size only, although the same maturity controls governed terminal
estimation and fallback in both forests. We used \(\varepsilon=10^{-4}\),
\(\lambda=0.05\), and implemented \(\texttt{min.gain}=0\). Hence the
rate-based stopping threshold in the asymptotic argument is not claimed for
this application. A terminal leaf failing support constraints reverted to its
nearest mature ancestor; fallback is the mean across trees of the unweighted
fraction of leaves requiring this reversion.

The single-tree displays used maximum depth two, a 20-point tail grid, and
\[
n_{\min}=\max\{80,\min(300,\lfloor0.02n\rfloor)\},\quad
r_8=\max\{30,\min(150,\lfloor0.01n\rfloor)\},
\]
\[
r_{10}=\max\{15,\min(60,\lfloor0.004n\rfloor)\},\quad
d_{<8}=\max\{10,\min(50,\lfloor0.005n\rfloor)\}.
\]
Marginal-KM censoring survival was truncated below at 0.02 for Brier
calculations; the censoring sensitivity used fivefold cross-fitted Cox models.
The paired validation bootstrap conditions on the fitted predictions, the
censoring fit, the original calibration bins, and the original extreme-group
assignments. It quantifies validation-cohort resampling uncertainty conditional
on model development, not uncertainty from refitting the complete pipeline.
Extreme groups were formed by stable ordering of predicted cure probability
and assigning exactly \(\lfloor0.2n\rfloor\) subjects to each disjoint tail.
The outcome contrast within those fixed groups was the held-out 8--10-year
survival estimate.

\subsection{Exploratory breadth across cancer sites}
\label{supp:fosp_breadth}

We used a deliberately exploratory, post-selected analysis to ask whether the
variance reduction from aggregation could translate into broader predictive
gains. All 35 cancer sites meeting the temporal-sample requirements were
screened with a 200-tree CBST forest. Twenty-one sites had both a larger
Harrell C-index and a smaller 8-year IPCW Brier score than a single
CBST. We then refitted the leading eligible primary sites using
the complete 2000--2009 training sample and three independently seeded
1000-tree CBST forests, with evaluation in 2010--2013. The single tree and
forest used the same cure-fraction based splitting criterion and tail-support rules; thus
this comparison isolates the effect of forest aggregation rather than the
choice of split target.

\begin{figure}[!htbp]
\centering
\includegraphics[width=\textwidth]{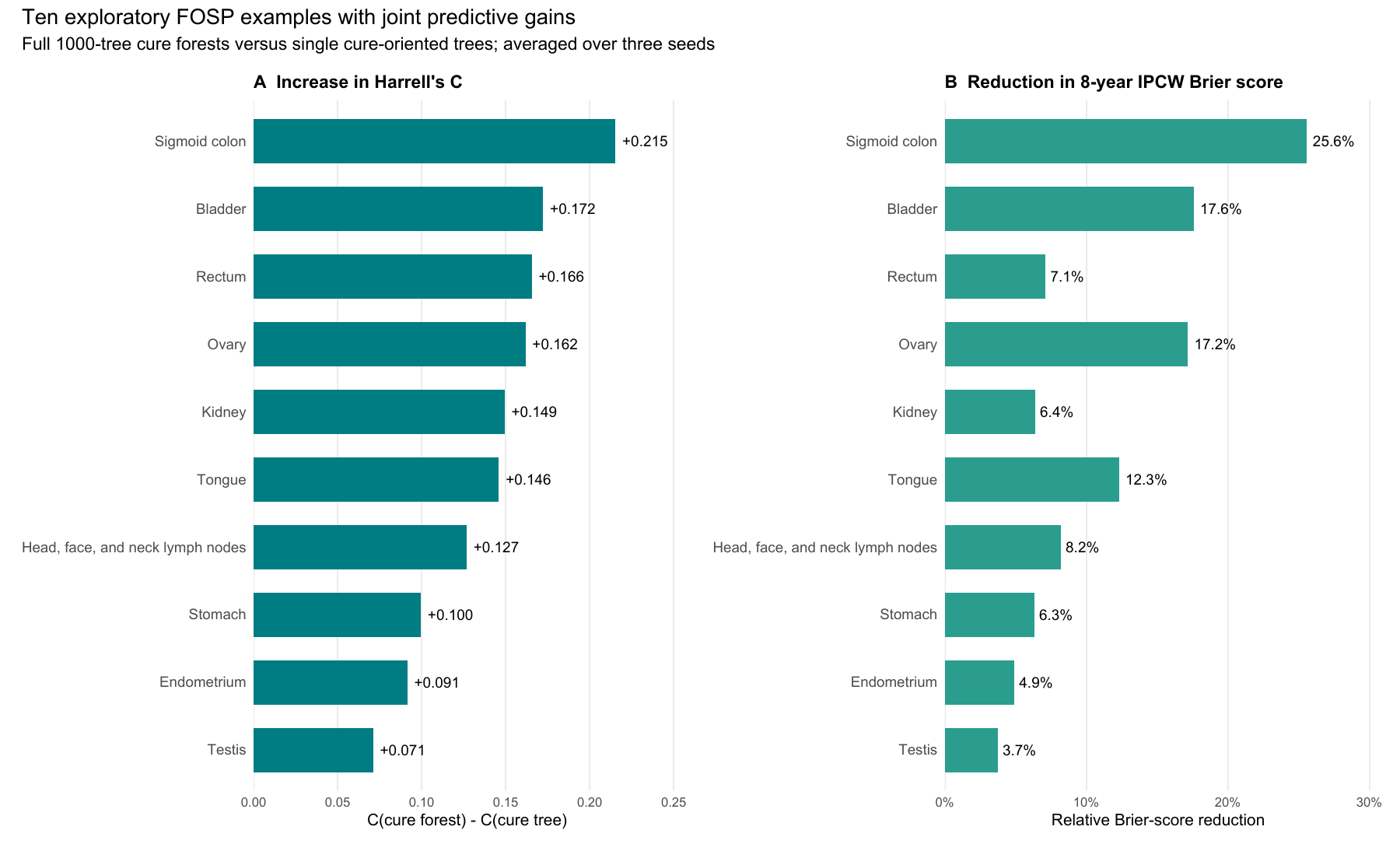}
\caption{Exploratory predictive gains for ten post-selected FOSP cancer sites.
Panel A shows the increase in Harrell C-index and Panel B the relative
reduction in 8-year IPCW Brier score for a 1000-tree CBST forest versus a
single CBST. Results are means over three forest seeds.}
\label{fig:supp_fosp_breadth}
\par\small\textit{Alt text:} Two horizontal bar charts show positive
forest-minus-tree C-index differences and positive relative Brier-score
reductions for sigmoid colon, testis, ovary, bladder, head, face, and neck
lymph nodes, kidney, rectum, tongue, stomach, and endometrial cancer.
\end{figure}

Supplementary Table~\ref{tab:supp_fosp_breadth} reports ten sites with a joint
mean gain after coherent tree-first aggregation. The C-index improvement ranged
from 0.071 to 0.215, and the relative reduction in 8-year Brier score ranged
from 3.7\% to 25.6\% (Supplementary
Figure~\ref{fig:supp_fosp_breadth}). The sites were chosen using these same
temporal-validation outcomes, so the display is post-selected and not
independent confirmation; no selection-adjusted uncertainty claim is made.

\begin{table}[!htbp]
\centering
\scriptsize
\caption{Exploratory temporal-validation examples with joint predictive gains.
Entries are means over three independently seeded 1000-tree CBST forests.
\(\Delta C=C_{\mathrm{forest}}-C_{\mathrm{tree}}\); Brier reduction is
\(100(B_{\mathrm{tree}}-B_{\mathrm{forest}})/B_{\mathrm{tree}}\) at 8 years.
Cancer sites were post-selected after a 200-tree screen.}
\label{tab:supp_fosp_breadth}
\resizebox{\textwidth}{!}{%
\begin{tabular}{lrrrrrrrr}
\toprule
Cancer site & Training & Validation & Tree \(C\) & Forest \(C\) &
\(\Delta C\) & Tree Brier & Forest Brier & Reduction (\%) \\
\midrule
Sigmoid colon & 12,972 & 9,153 & 0.520 & 0.735 & 0.215 & 0.251 & 0.187 & 25.6 \\
Testis        & 1,659  & 1,091 & 0.669 & 0.740 & 0.071 & 0.106 & 0.102 & 3.7 \\
Ovary         & 4,953  & 2,617 & 0.540 & 0.702 & 0.162 & 0.245 & 0.203 & 17.2 \\
Bladder       & 6,700  & 4,355 & 0.581 & 0.754 & 0.172 & 0.223 & 0.184 & 17.6 \\
Head, face, and neck lymph nodes
              & 7,988  & 4,541 & 0.537 & 0.664 & 0.127 & 0.227 & 0.209 & 8.2 \\
Kidney        & 4,420  & 3,850 & 0.554 & 0.703 & 0.149 & 0.227 & 0.213 & 6.4 \\
Rectum        & 10,579 & 7,255 & 0.523 & 0.688 & 0.166 & 0.250 & 0.233 & 7.1 \\
Tongue        & 3,415  & 1,563 & 0.531 & 0.677 & 0.146 & 0.239 & 0.210 & 12.3 \\
Stomach       & 15,929 & 9,013 & 0.554 & 0.654 & 0.100 & 0.195 & 0.183 & 6.3 \\
Endometrium   & 5,348  & 3,442 & 0.602 & 0.693 & 0.091 & 0.195 & 0.186 & 4.9 \\
\bottomrule
\end{tabular}}
\end{table}

\subsection{Additional construction and reporting details}
\label{supp:construction_details}

\paragraph{Cross-fitted IPCW nodes.}
For covariate-dependent censoring, let
\(G_C(t\mid x)=P(C>t\mid X=x)\), fix folds
\(\mathcal I_1,\ldots,\mathcal I_K\), and estimate
\(\widehat G_C^{(-k)}\) without fold \(k\). The node survival estimator used for
observation \(i\in\mathcal I_k\) is
\[
\widehat S_A^{\mathrm{IPCW}}(t)
=
\frac{1}{n_A}
\sum_{k=1}^K\sum_{i\in\mathcal I_k:X_i\in A}
\frac{I(Y_i>t)}{\widehat G_C^{(-k)}(t\mid X_i)}.
\]
The implementation truncates \(\widehat G_C\) below by a prespecified
positivity constant. This is a Horvitz--Thompson estimator and need not be
monotone. For the scalar tail estimate, the implementation first averages the
unprojected Horvitz--Thompson values over the tail grid and node, then projects
that scalar to \([0,1]\). Finite-time terminal survival curves are clipped
pointwise to \([0,1]\). A marginal Kaplan--Meier censoring model, a cross-fitted
Cox model, or externally
estimated censoring probabilities may be supplied. If the nuisance is fitted
once using the full sample and shared by all trees, ordinary tree-kernel IJ
theory omits its additional influence contribution.

\paragraph{Variance normalization.}
For a KM tail grid \(t_1<\cdots<t_J\), weights \(w_j\), and cumulative
Greenwood factor
\[
\widehat G_A(t)=\sum_{u\le t}
\frac{d_A(u)}{Y_A(u)\{Y_A(u)-d_A(u)\}},
\]
the implementation retains covariance across tail times:
\[
\widehat V\{\widehat\pi_A^{\mathrm{KM}}\}
=\sum_{j=1}^J\sum_{k=1}^J
w_j w_k\widehat S_A(t_j)\widehat S_A(t_k)
\widehat G_A(t_j\wedge t_k).
\]
Equivalently, this is the sum over Greenwood jumps of the squared weighted
survival remaining after each jump. For IPCW contributions
\(h_i=J^{-1}\sum_j I(Y_i>t_j)/\widehat G_C(t_j\mid X_i)\), the node normalizer
is \(s_A^2(h)/n_A\). With
\(p_A=\min\{1-\varepsilon,\max(\varepsilon,\widehat\pi_A)\}\), the logit-scale
normalizer is
\[
\widehat V\{\logit(\widetilde\pi_A)\}
=\frac{\widehat V(\widehat\pi_A)}{p_A^2(1-p_A)^2}.
\]
Clipping is numerical regularization and does not resolve boundary
nonregularity. The balance factor protects against small unstable daughters,
and \(\lambda\) prevents an almost-zero normalizer from dominating split
ranking.

\paragraph{Optional scores.}
When a single plateau is unstable, an integrated alternative is
\[
G_{\mathrm{tail}}(s;A)=\omega_s
\int_{\mathcal T_A}
\{\widehat S_{A_L(s)}(t)-\widehat S_{A_R(s)}(t)\}^2\,d\Omega_A(t).
\]
When both incidence and latency are primary, a hybrid is
\[
G_\alpha(s;A)=\alpha G_{\mathrm{cure}}(s;A)
+(1-\alpha)G_{\mathrm{latency}}(s;A),\qquad 0\le\alpha\le1.
\]
The main theory and experiments concern \(G_{\mathrm{cure}}\). These extensions
require their own population targets, uniform score laws, separation
conditions, and, for posterior-weighted latency scores, a specified
nuisance-estimation scheme.

\paragraph{Complete fitting workflow.}
Required inputs are observed time, event indicator, covariates, a prespecified
or restricted mature-tail rule, KM or IPCW node estimation, and the usual
forest controls. At each node, a random variable subset and candidate
cutpoints are formed. For the cure-fraction based splitting criterion, daughters violating size,
event-count, risk-set, or censoring-support constraints are removed. Log-rank
candidate scoring enforces minimum daughter size only; the cure-specific
maturity controls still enter terminal estimation and fallback. Remaining cure
curves are evaluated on a common grid. The split score uses the stabilized estimate
\(\widetilde\pi_A\), whereas the terminal prediction uses the
support-aware \(\overline\pi_A\). After tree growth, susceptible survival is
reconstructed by the mixture identity. No additional EM refinement is part of
the method.

For predictive fitting, out-of-bag survival, tail support, and stability over a
prespecified grid may guide tuning. Latent cure MSE and cure-status Brier score
are available only in simulation and cannot tune real-data fits. For formal
inference, all controls must instead be fixed or selected on independent data
unless the selection step is incorporated into the analysis. An inferential
fit stores tree predictions and inclusion indicators, applies the fixed-size
without-replacement factor, and forms paired contrasts within tree. The
finite-\(B\) subtraction remains diagnostic. Reported output should include the
exact tail rule, maturity thresholds, effective risk sets, and the frequency
with which terminal leaves revert to a mature ancestor.

For a contrast, tree predictions and IJ influences are differenced within the
same tree before the finite-\(B\) calculation, retaining their pairing. When
requested, censoring nuisances use observation-level cross-fitting; a nuisance
estimated once on all training observations is shared across trees and falls
under the inferential restriction stated above.

\paragraph{Coherent forest aggregation.}
For tree-specific cure estimates \(\widehat\pi_b(x)\) and identifiable
susceptible curves \(\widehat S_{u,b}(t\mid x)\), population survival is
\begin{equation}
\widehat S_{\mathrm{CBST}}(t\mid x)
=\frac{1}{B}\sum_{b=1}^B
\left[\widehat\pi_b(x)+
\{1-\widehat\pi_b(x)\}\widehat S_{u,b}(t\mid x)\right].
\label{eq:forest_population_survival}
\end{equation}
When \(\widehat\pi_{\mathrm{CBST}}(x)<1\), the forest-level susceptible curve
is
\begin{equation}
\widehat S_{u,\mathrm{CBST}}^{\mathrm{coh}}(t\mid x)
=
\frac{\widehat S_{\mathrm{CBST}}(t\mid x)
-\widehat\pi_{\mathrm{CBST}}(x)}
{1-\widehat\pi_{\mathrm{CBST}}(x)}.
\label{eq:coherent_forest_latency}
\end{equation}
This is a weighted average of valid tree-level susceptible curves, so the
mixture identity is preserved without forest-level truncation. The
implementation first replaces each raw tree survival by its coherent mixture
curve and only then averages. All finite-time results reported here use this
tree-first order. A tree with cure estimate one contributes zero
weighted susceptible survival; its unweighted susceptible curve is not treated
as identified.


\begin{thebibliography}{}

\bibitem[Amico and Van~Keilegom, 2018]{AmicoVanKeilegom2018}
Amico, M. and Van~Keilegom, I. (2018).
\newblock Cure models in survival analysis.
\newblock {\em Annual Review of Statistics and Its Application}, 5:311--342.

\bibitem[Aselisewine and Pal, 2023]{aselisewine2023}
Aselisewine, W. and Pal, S. (2023).
\newblock On the integration of decision trees with mixture cure model.
\newblock {\em Statistics in Medicine}, 42(23):4111--4127.

\bibitem[Aselisewine and Pal, 2024]{aselisewine2024}
Aselisewine, W. and Pal, S. (2024).
\newblock Enhancing cure rate analysis through integration of machine learning
  models: A comparative study.
\newblock {\em Statistics and Computing}, 34(4):142.

\bibitem[Athey et~al., 2019]{AtheyTibshiraniWager2019}
Athey, S., Tibshirani, J., and Wager, S. (2019).
\newblock Generalized random forests.
\newblock {\em The Annals of Statistics}, 47(2):1148--1178.

\bibitem[Berkson and Gage, 1952]{berkson1952}
Berkson, J. and Gage, R.~P. (1952).
\newblock Survival curve for cancer patients following treatment.
\newblock {\em Journal of the American Statistical Association},
  47(259):501--515.

\bibitem[Boag, 1949]{boag1949}
Boag, J.~W. (1949).
\newblock Maximum likelihood estimates of the proportion of patients cured by
  cancer therapy.
\newblock {\em Journal of the Royal Statistical Society: Series B
  (Methodological)}, 11(1):15--44.

\bibitem[Breiman, 1996]{Breiman1996Bagging}
Breiman, L. (1996).
\newblock Bagging predictors.
\newblock {\em Machine Learning}, 24(2):123--140.

\bibitem[Breiman, 2001]{breiman2001}
Breiman, L. (2001).
\newblock Random forests.
\newblock {\em Machine Learning}, 45:5--32.

\bibitem[Breiman et~al., 1984]{BreimanEtAl1984CART}
Breiman, L., Friedman, J.~H., Olshen, R.~A., and Stone, C.~J. (1984).
\newblock {\em Classification and Regression Trees}.
\newblock Wadsworth International Group, Belmont, CA.

\bibitem[Chen et~al., 1999]{chen1999}
Chen, M.-H., Ibrahim, J.~G., and Sinha, D. (1999).
\newblock A new bayesian model for survival data with a surviving fraction.
\newblock {\em Journal of the American Statistical Association},
  94(447):909--919.

\bibitem[Chiou et~al., 2025]{ChiouEtAl2025}
Chiou, S.~H., Su, C.-L., and Lin, F.-C. (2025).
\newblock Fitting cure models using pseudo-observations approaches with the {R}
  package pseudocure.
\newblock {\em Japanese Journal of Statistics and Data Science}, 8:603--639.

\bibitem[Cui et~al., 2022]{CuiEtAl2022}
Cui, Y., Zhu, R., Zhou, M., and Kosorok, M.~R. (2022).
\newblock Consistency of survival tree and forest models: Splitting bias and
  correction.
\newblock {\em Statistica Sinica}, 32(3):1245--1267.

\bibitem[de~Almeida et~al., 2022]{almeida2022fosp}
de~Almeida, R.~J., de~Moraes~Luizaga, C.~T., Eluf-Neto, J., de~Carvalho~Nunes,
  H.~R., Pessoa, E.~C., and Murta-Nascimento, C. (2022).
\newblock Impact of educational level and travel burden on breast cancer stage
  at diagnosis in the state of sao paulo, brazil.
\newblock {\em Scientific Reports}, 12:8357.

\bibitem[Escobar-Bach and Van~Keilegom, 2019]{EscobarBachVanKeilegom2019}
Escobar-Bach, M. and Van~Keilegom, I. (2019).
\newblock Non-parametric cure rate estimation under insufficient follow-up by
  using extremes.
\newblock {\em Journal of the Royal Statistical Society: Series B (Statistical
  Methodology)}, 81(5):861--880.

\bibitem[Ezquerro et~al., 2023]{ezquerro2023}
Ezquerro, A., Cancela, B., and L{\'o}pez-Cheda, A. (2023).
\newblock On the reliability of machine learning models for survival analysis
  when cure is a possibility.
\newblock {\em Mathematics}, 11(19):4150.

\bibitem[Farewell, 1982]{farewell1982}
Farewell, V.~T. (1982).
\newblock The use of mixture models for the analysis of survival data with
  long-term survivors.
\newblock {\em Biometrics}, 38(4):1041--1046.

\bibitem[{Funda\c{c}\~ao Oncocentro de S\~ao Paulo}, 2026]{FOSP2026}
{Funda\c{c}\~ao Oncocentro de S\~ao Paulo} (2026).
\newblock Hospital cancer registry of the state of s\~ao paulo ({RHC/SP}).
\newblock FOSP public data archive.
\newblock Data snapshot downloaded May 27, 2026.

\bibitem[Han et~al., 2024]{HanEtAl2024}
Han, B., Wang, X., and Sun, L. (2024).
\newblock Large-scale survival analysis with a cure fraction.
\newblock {\em Biometrics}, 80(4):ujae138.

\bibitem[Ishwaran et~al., 2008]{ishwaran2008}
Ishwaran, H., Kogalur, U.~B., Blackstone, E.~H., and Lauer, M.~S. (2008).
\newblock Random survival forests.
\newblock {\em The Annals of Applied Statistics}, 2(3):841--860.

\bibitem[Jiang et~al., 2019]{jiang2019}
Jiang, C., Wang, Z., and Zhao, H. (2019).
\newblock A prediction-driven mixture cure model and its application in credit
  scoring.
\newblock {\em European Journal of Operational Research}, 277(1):20--31.

\bibitem[Kaplan and Meier, 1958]{KaplanMeier1958}
Kaplan, E.~L. and Meier, P. (1958).
\newblock Nonparametric estimation from incomplete observations.
\newblock {\em Journal of the American Statistical Association},
  53(282):457--481.

\bibitem[Kuk and Chen, 1992]{kuk1992}
Kuk, A. Y.~C. and Chen, C.-H. (1992).
\newblock A mixture model combining logistic regression with proportional
  hazards regression.
\newblock {\em Biometrika}, 79(3):531--541.

\bibitem[LeBlanc and Crowley, 1993]{LeBlancCrowley1993}
LeBlanc, M. and Crowley, J. (1993).
\newblock Survival trees by goodness of split.
\newblock {\em Journal of the American Statistical Association},
  88(422):457--467.

\bibitem[Linero et~al., 2026]{LineroRubioBasak2026BartCure}
Linero, A.~R., Rubio, F.~J., and Basak, P. (2026).
\newblock Bayesian causal machine learning for cure models.
\newblock {\em arXiv preprint arXiv:2606.11405}.

\bibitem[L{\'o}pez-Cheda et~al., 2021]{LopezChedaEtAl2021Npcure}
L{\'o}pez-Cheda, A., J{\'a}come, M.~A., and L{\'o}pez-de Ullibarri, I. (2021).
\newblock npcure: An {R} package for nonparametric inference in mixture cure
  models.
\newblock {\em The R Journal}, 13(1):21--41.

\bibitem[Maller and Zhou, 1996]{maller1996}
Maller, R.~A. and Zhou, X. (1996).
\newblock {\em Survival Analysis with Long-Term Survivors}.
\newblock John Wiley \& Sons, Chichester.

\bibitem[Mentch and Hooker, 2016]{MentchHooker2016}
Mentch, L. and Hooker, G. (2016).
\newblock Quantifying uncertainty in random forests via confidence intervals
  and hypothesis tests.
\newblock {\em Journal of Machine Learning Research}, 17(26):1--41.

\bibitem[Overgaard et~al., 2019]{OvergaardEtAl2019}
Overgaard, M., Parner, E.~T., and Pedersen, J. (2019).
\newblock Pseudo-observations under covariate-dependent censoring.
\newblock {\em Journal of Statistical Planning and Inference}, 202:112--122.

\bibitem[Pal and Aselisewine, 2023]{pal2023}
Pal, S. and Aselisewine, W. (2023).
\newblock A semiparametric promotion time cure model with support vector
  machine.
\newblock {\em The Annals of Applied Statistics}, 17(3):2680--2699.

\bibitem[Peng and Dear, 2000]{peng2000}
Peng, Y. and Dear, K. B.~G. (2000).
\newblock A nonparametric mixture model for cure rate estimation.
\newblock {\em Biometrics}, 56(1):237--243.

\bibitem[Rodrigues et~al., 2009]{rodrigues2009}
Rodrigues, J., Cancho, V.~G., de~Castro, M., and Louzada-Neto, F. (2009).
\newblock On the unification of long-term survival models.
\newblock {\em Statistics \& Probability Letters}, 79(6):753--759.

\bibitem[Sy and Taylor, 2000]{sy2000}
Sy, J.~P. and Taylor, J. M.~G. (2000).
\newblock Estimation in a cox proportional hazards cure model.
\newblock {\em Biometrics}, 56(1):227--236.

\bibitem[Tsodikov et~al., 2003]{tsodikov2003}
Tsodikov, A.~D., Ibrahim, J.~G., and Yakovlev, A.~Y. (2003).
\newblock Estimating cure rates from survival data: An alternative to
  two-component mixture models.
\newblock {\em Journal of the American Statistical Association},
  98(464):1063--1078.

\bibitem[Wager and Athey, 2018]{WagerAthey2018}
Wager, S. and Athey, S. (2018).
\newblock Estimation and inference of heterogeneous treatment effects using
  random forests.
\newblock {\em Journal of the American Statistical Association},
  113(523):1228--1242.

\bibitem[Wager et~al., 2014]{WagerHastieEfron2014}
Wager, S., Hastie, T., and Efron, B. (2014).
\newblock Confidence intervals for random forests: The jackknife and the
  infinitesimal jackknife.
\newblock {\em Journal of Machine Learning Research}, 15(48):1625--1651.

\bibitem[Yakovlev and Tsodikov, 1996]{yakovlev1996}
Yakovlev, A.~Y. and Tsodikov, A.~D. (1996).
\newblock {\em Stochastic Models of Tumor Latency and Their Biostatistical
  Applications}.
\newblock World Scientific, Singapore.

\bibitem[Yuen and Musta, 2026]{YuenMusta2026}
Yuen, T.~P. and Musta, E. (2026).
\newblock Testing for sufficient follow-up in survival data with a cure
  fraction.
\newblock {\em Biometrical Journal}, 68(2):e70121.

\end{thebibliography}
\end{document}